\pdfoutput=1
\documentclass[10pt, twocolumn, comsoc]{IEEEtran}

\usepackage{graphicx,epsfig}
\usepackage[noadjust]{cite}
\usepackage{mcite}
\usepackage{amsfonts,helvet}
\usepackage{fancyhdr}
\usepackage{threeparttable}
\usepackage{epsf,epsfig}
\usepackage{amsthm}
\usepackage{amsmath}
\usepackage{siunitx}
\usepackage{amssymb}
\usepackage{stfloats}
\usepackage{comment}

\usepackage{tabularx}

\usepackage[colorlinks=true, linkcolor=blue]{hyperref}
\usepackage{cleveref}

\usepackage{dsfont}
\usepackage{subfigure}
\usepackage{color}
\usepackage{enumerate}
\usepackage{gensymb}
\usepackage{cancel}
\usepackage{lipsum}
\usepackage{mathtools}
\usepackage{cuted}
\usepackage{bbm}
\usepackage[linesnumbered,ruled]{algorithm2e}

\newtheorem{remark}{Remark}

\usepackage{eucal}

\newcommand{\IEEEpostingnotice}{}
\usepackage{eso-pic}
\AddToShipoutPictureFG*{\AtPageUpperLeft{\put(0,-22){\makebox[\paperwidth][c]{%
  \parbox[t]{0.86\textwidth}{\centering\footnotesize\IEEEpostingnotice}}}}}

\begin{document}

\title{Learning MMSE Filters for OFDM Channel Estimation: Attention Transformer Gains at Linear Inference}

\author{TaeJun Ha, Chaehyun Jung, Hyeonuk Kim, Jeongwoo Park, and Jeonghun Park

\thanks{This work was supported in part by the 6GARROW project funded by the Smart Networks and Services Joint Undertaking (SNS JU) under the European Union's Horizon Europe research and innovation programme (Grant Agreement No.~101192194) and by an IITP grant funded by the Korean government (MSIT) (No.~RS-2024-00435652); in part by an IITP grant funded by MSIT (No.~RS-2024-00395824, ``Development of Cloud-virtualized RAN (vRAN) system supporting upper-midband''); and in part by the IITP Digital Innovation Talent Short-Term Intensive Program grant funded by MSIT (No.~RS-2024-00404972).

The authors are with the School of Electrical and Electronic Engineering, Yonsei University, South Korea (e-mail: {\texttt{tjha@yonsei.ac.kr, jch0624@yonsei.ac.kr, garksi11@yonsei.ac.kr, simyffff@yonsei.ac.kr, jhpark@yonsei.ac.kr}}).
}
}

\maketitle \setcounter{page}{1}
\begin{abstract}
    In orthogonal frequency division multiplexing (OFDM), accurate channel estimation is crucial. Classical signal processing-based approaches, such as linear minimum mean-squared error (LMMSE) estimation, often require second-order statistics that are difficult to obtain in practice. Recent deep neural network (DNN)-based methods have been introduced to address this, but they often suffer from high inference complexity.
    This paper proposes an Attention-aided MMSE (A-MMSE), a model-based DNN framework that learns the linear MMSE filter via the Attention Transformer. Once trained, the A-MMSE performs channel estimation through a single linear operation, eliminating nonlinear activations during inference and thus reducing computational complexity. To improve the learning efficiency of the A-MMSE, we develop a two-stage Attention encoder that captures the frequency and temporal correlation structure of OFDM channels.
    We also introduce a rank-adaptive extension that adjusts the filter rank at deployment time, enabling efficient operation under resource-constrained receivers.
    Numerical simulations show that A-MMSE consistently outperforms baseline methods across a wide range of signal-to-noise ratio (SNR) conditions. In particular, the A-MMSE and its rank-adaptive extension provide an improved performance-complexity trade-off. %
\end{abstract}


\section{Introduction}
Orthogonal frequency division multiplexing (OFDM) is a standard waveform in modern wireless communication systems. OFDM was initially adopted in earlier standards such as Wi-Fi (IEEE 802.11) and WiMAX, and was later introduced into the 3GPP-based cellular ecosystem with LTE. This adoption paved the way for OFDM's continued prominence in 5G \cite{shafi:jsac:17, guan:jsac:17} and its expected role in future 6G technologies \cite{liy:twc:21}.

One of the critical problems in OFDM is channel estimation, which is essential for enabling coherent detection.
The goal of OFDM channel estimation is to accurately estimate the channel state information for each subcarrier using noisy pilot symbols.
Classical signal processing (SP)-based methods for OFDM channel estimation include least squares (LS) and linear minimum mean-squared error (LMMSE) estimation~\cite{edfors:tcom:98, li:tvt:00}.
LS is simple to implement, but its performance is limited because it cannot incorporate prior OFDM channel statistics.
LMMSE, in contrast, requires second-order statistics, such as noise variance and correlation across subcarriers, which are difficult to obtain accurately in practical scenarios \cite{liu:survey:14}.

Recently, driven by the success of deep neural networks (DNNs) in signal processing, DNN-aided OFDM channel estimation has gained considerable attention.
DNNs can capture complex data relationships without explicit prior statistical knowledge, and several prior works have explored this for OFDM channel estimation.
In~\cite{soltani:commlett:19}, ChannelNet treats the OFDM channel as a 2D image, applying a super-resolution network followed by a denoising CNN.
In \cite{li:wcl:20}, using a similar modeling approach to \cite{soltani:commlett:19},
ReEsNet was proposed, where a residual learning-based DNN was used to reconstruct the OFDM channel. This method achieves lower computational cost and higher accuracy than \cite{soltani:commlett:19}.
In \cite{luan:wsa:21}, a residual convolutional network combined with bilinear interpolation was proposed for OFDM channel estimation, offering enhanced flexibility to varying pilot patterns and reduced complexity.
Beyond deterministic models, \cite{balevi:jsac:21} optimized a deep generative model using compressive pilot measurements for high-dimensional channel estimation.
This method leverages learned priors to reduce pilot overhead and surpass traditional sparse recovery techniques.

\begin{figure*}[t]
    \centering
    \includegraphics[width=1\textwidth]{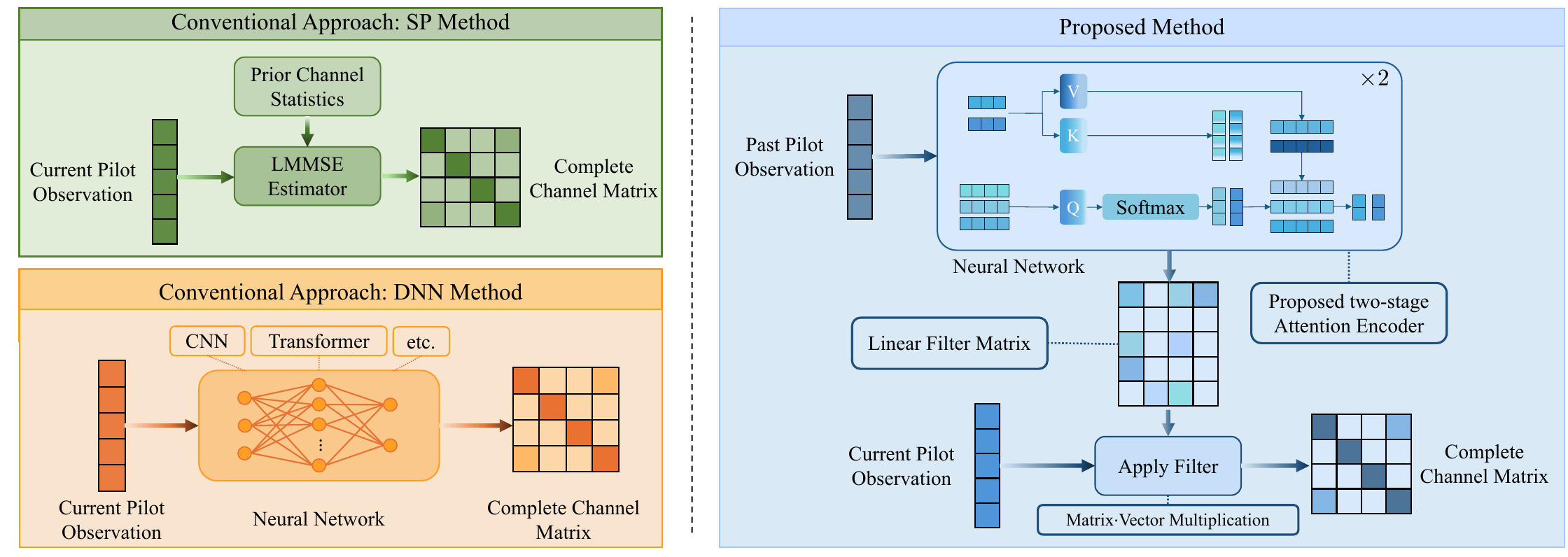}
    \caption{Comparison of OFDM channel estimation structure: conventional approaches vs. the proposed A-MMSE approach.}
    \label{fig:comparison}
\end{figure*}

In particular, with the advent of the Attention Transformer architecture \cite{vaswani2017attention}, some recent works investigated Transformer models as a compelling solution for wireless channel estimation problems, owing to their representational capacity.
In \cite{li2023wireless}, super-resolution techniques were integrated with an Attention-based encoder-decoder framework to enhance OFDM channel estimation accuracy.
In \cite{luan2023channelformer}, Channelformer was developed, where an Attention-based encoder-decoder neural architecture was designed to incorporate input precoding.
A common structure in the prior studies using DNNs for channel estimation  \cite{soltani:commlett:19, li:wcl:20, luan:wsa:21, balevi:jsac:21, li2023wireless, luan2023channelformer} is the use of a pre-trained neural network that directly maps noisy pilot signals to the estimated channel matrix, as illustrated in the left panel of Fig.~\ref{fig:comparison}.

Despite their significant advantages over classical SP-based channel estimation methods, existing DNN-based approaches have several limitations. We summarize these as follows.
\begin{itemize}
    \item {\textbf{Computational overhead}}:
Typically, DNN-based models involve a large number of parameters (approximately $53{,}000$ for ReEsNet \cite{li:wcl:20} and $110{,}000$ for Channelformer \cite{luan2023channelformer}),
resulting in significant computational and memory overhead.
Furthermore, reliance on nonlinear activations (e.g., ReLU~\cite{nair2010rectified}, tanh~\cite{lecun2012efficient}, ELU~\cite{clevert2015fast}, and pooling~\cite{scherer2010evaluation}) limits hardware acceleration and complicates practical deployment.

\item {\textbf{Limited interpretability}}: Existing DNN-based approaches are typically applied directly to OFDM channel estimation in a purely data-driven manner, without incorporating domain knowledge derived from mathematical models. However, studies such as \cite{greenberg:neurips:23, ngu:wcommagg:23} showed that integrating domain-specific insights into neural network design can lead to notable performance gains over purely data-driven methods.

\item {\textbf{Lack of flexibility}}:
For practical deployment, it is often desirable for channel estimation methods to adaptively adjust computational load.
For example, a user equipment (UE) with constrained processing resources should be able to reduce estimator complexity by adjusting the number of parameters.
However, most existing DNN-based approaches lack this adaptability, as they provide no explicit mechanisms to control the model size or computational cost within the DNN architecture.

\end{itemize}
We note that these limitations primarily arise from employing DNNs as end-to-end channel estimators, where the model learns a direct mapping from noisy pilot signals to channel estimates solely based on data.

To address the challenges of purely data-driven approaches, recent research has actively explored the concept of model-based DNNs \cite{mbdnn:book:23}.
The core idea is to leverage a hybrid method that combines classical SP-based approaches and data-driven DNN-based approaches, thereby gaining mathematical interpretability without sacrificing model expressiveness.
For instance, KalmanNet~\cite{kalmannet:tsp:22} retains the recursive update structure of the Kalman filter~\cite{welch:kalman} while learning the Kalman gain via a Long Short-Term Memory (LSTM) \cite{lstm:97} or gated recurrent units (GRU) \cite{cho:gru:14}.
Thanks to its hybrid design, KalmanNet \cite{kalmannet:tsp:22} effectively addresses state estimation with partially known models, and has since inspired several extensions \cite{choi:tvt:23, shen2025kalmanformer, aikalman:25, jungBussgang}.
In parallel, model-based DNNs have been applied to multiple-input multiple-output (MIMO) beamforming \cite{xu:twc:25}, MIMO detection \cite{ngu:wcommagg:23, khani:twc:20},
and IRS/RIS-assisted channel estimation, where model-based single-layer DNN architectures  recover channel quantities from reference signal received power (RSRP) measurements via a compact low-rank linear mapping~\cite{Sun2025twc, Sun2024twc, Liu2025CL}.
A seemingly straightforward route to OFDM is to apply such model-based DNNs in a per-subcarrier manner, but this discards the cross-subcarrier and cross-symbol correlations central to OFDM and amounts to merely a block-diagonal restriction of the joint frequency--temporal filter pursued in this work.
Beyond this restricted instance, extending model-based DNNs to OFDM in a non-trivial way poses three intertwined challenges: learning the high-dimensional joint filter without overfitting; handling non-stationary statistics under high-mobility conditions; and exploiting the Kronecker-separable covariance structure~\cite{Yoojin20152DLMMSE}, $\mathbf{R}_{\mathrm{full}} \approx \mathbf{R}_f \otimes \mathbf{R}_t$, through a domain-specific architecture rather than a generic neural network.
In contrast to the compact single-layer mapping pursued in~\cite{Sun2025twc, Sun2024twc, Liu2025CL}, the OFDM linear estimation filter spans a high-dimensional joint frequency-temporal space.
To our knowledge, no prior model-based DNN framework addresses these three challenges in an efficient manner while achieving high estimation accuracy. We fill this gap by introducing a two-stage Attention encoder that exploits the covariance structure as an inductive bias, combining the learning capability of DNNs with domain knowledge from classical SP-based linear estimation.

In this paper, we propose an Attention-aided MMSE (A-MMSE) method for OFDM channel estimation.
The key idea is to use the Attention Transformer to learn linear channel estimation filter coefficients, rather than channel estimates directly.
The fundamental difference between conventional DNN-based approaches and the proposed method is illustrated in Fig.~\ref{fig:comparison}.
Once trained, the proposed A-MMSE performs channel estimation through a single linear operation, decoupling inference from the learning procedure.
The A-MMSE is further extended with a rank-adaptive capability, referred to as the rank-adaptive A-MMSE (RA-A-MMSE).
The rank of the RA-A-MMSE filter matrix is adjustable at deployment time, enabling computational overhead reduction with minimal loss in estimation accuracy.

We further develop a two-stage Attention encoder to enhance estimation performance.
Specifically, we first observe that the OFDM channel's correlation matrix can be effectively decomposed into frequency and temporal domain correlations.
Building on this insight, our two-stage Attention encoder consists of Frequency Encoder and Temporal Encoder, each designed to capture the corresponding correlation structure.
The Frequency Encoder applies Multi-Head Self-Attention with a linear embedding whose dimension equals the number of subcarriers.
Its output is projected to the Temporal Encoder, which applies Multi-Head Self-Attention with an embedding dimension equal to the total number of OFDM resource elements.
By sequentially integrating the Frequency and Temporal Attention encoder blocks, our method effectively reflects complementary relationships across the frequency and temporal domains, extracting highly relevant latent features.
These refined features are then decoded by a residual fully connected network, precisely mapping them to the linear filter coefficients of the proposed A-MMSE method.
Thanks to the two-stage Attention structure, OFDM channels' inherent correlation characteristics are effectively incorporated in a form particularly suitable for A-MMSE filter design, significantly enhancing the overall estimation performance.

Numerical simulations with the COST2100 models~\cite{COST2100} demonstrate that A-MMSE outperforms both SP-based estimators (e.g., LS and LMMSE) and DNN-based methods (e.g., ChannelNet \cite{soltani:commlett:19} and Channelformer \cite{luan2023channelformer}) across all evaluated conditions.
The RA-A-MMSE maintains comparable estimation accuracy while substantially reducing computational complexity.
In particular, averaged across all evaluated SNR conditions and scenarios, A-MMSE achieves approximately $61\%$ lower normalized mean-squared error (NMSE) than conventional LMMSE and $67\%$ lower NMSE than Channelformer. The gains stem from the two-stage encoder, where frequency-domain features are extracted first and then refined through temporal correlation modeling.
Ablation studies further validate that the two encoder stages are complementary and indispensable, where removing either stage leads to substantial NMSE degradation, and the temporal encoder becomes particularly critical under high-mobility conditions.
Moreover, the RA-A-MMSE further reduces computational complexity through adaptive rank reduction, while retaining estimation accuracy comparable to the full-rank A-MMSE. At $10\%$ rank, the RA-A-MMSE requires approximately $10\%$ of the FLOPs of LMMSE, while achieving approximately $51\%$ lower NMSE compared to Channelformer. The proposed methods also exhibit strong robustness to signal-to-noise ratio (SNR) mismatch, retaining performance across a wide range of unseen SNR conditions. This underscores the practical efficiency and suitability of the proposed method for real-world 5G/6G implementations.

The main contributions of this paper are as follows.
First, A-MMSE incorporates a two-stage Attention encoder that captures the frequency- and temporal-domain correlation structure of OFDM channels.
The encoder exploits the Kronecker-separable covariance structure as an inductive bias.
Averaged across all evaluated SNR conditions and scenarios, A-MMSE achieves approximately $61\%$ and $67\%$ lower NMSE than LMMSE and Channelformer, respectively.
Second, once trained, A-MMSE performs channel estimation via a single matrix-vector multiplication. Nonlinear activations are absent at inference, reducing computational complexity relative to conventional DNN-based estimators.
Third, the RA-A-MMSE adjusts the rank of the learned filter at deployment time. At $10\%$ rank, computational complexity is reduced by approximately $90\%$ relative to the full-rank A-MMSE, with minimal degradation in estimation accuracy.

The rest of the paper is organized as follows. Section~\ref{sec:sys} describes the system model, and Section~\ref{sec:conven} reviews the conventional SP-based and DNN-based channel estimators.
Section~\ref{sec:ammse} introduces the proposed A-MMSE method, and Section~\ref{sec:ra_ammse} further extends it to the RA-A-MMSE.
Numerical results are provided in Section~\ref{sec:sim}, and Section~\ref{sec:con} concludes the paper.

\textbf{Notations}: Bold lowercase and uppercase letters denote vectors and matrices, respectively.
We denote transpose by $(\cdot)^{\mathsf T}$, Hermitian transpose by $(\cdot)^{\mathsf H}$, column-wise vectorization by $\operatorname{vec}(\cdot)$, real and imaginary parts of a complex quantity by $\Re(\cdot)$ and $\Im(\cdot)$, the magnitude by $|\cdot|$, the Euclidean norm of a vector and the spectral norm of a matrix by $\|\cdot\|_2$, and the Frobenius norm by $\|\cdot\|_F$.
Element-wise (Hadamard) multiplication is denoted by $\circ$, the Kronecker product by $\otimes$, and statistical expectation is written as $\mathbb{E}[\cdot]$. The set difference is denoted by $\setminus$, horizontal (row-wise) stacking by $[\cdot,\cdot]$, and vertical (column-wise) stacking by $[\cdot;\cdot]$.

\section{System Model} \label{sec:sys}

We consider a single OFDM slot with $M$ OFDM symbols, each comprising $N$ subcarriers, for a total of $NM$ resource elements. The channel's frequency response corresponding to the $n$-th subcarrier and the $m$-th symbol is denoted as $H[n,m]$.
This work considers single-antenna OFDM, where $n$ and $m$ index the subcarrier and OFDM symbol, respectively.
Assuming that the cyclic-prefix (CP) is removed, the received signal model in the frequency-domain is given by
\begin{align}
    Y[n,m] = H[n,m] X[n,m] + Z[n,m], \; (n, m) \in \CMcal{T},\label{eq:Channel model}
\end{align}
where $X[n,m]$ is the transmitted signal, $Z[n,m]$ is the additive Gaussian noise, and $\CMcal{T}$ denotes the set of all resource elements.
We denote by \( \CMcal{P} \) the set of pilot indices, where \( |\CMcal{P}| = L \). This indicates that a total of \( L \) pilot symbols are used for channel estimation.
Accordingly, \( X[n,m] \) is known to the receiver for \( (n, m) \in \CMcal{P} \). Let \( \CMcal{D} \) be the set of indices corresponding to the data signals, such that \( \CMcal{P} \cup \CMcal{D} = \CMcal{T} \).
For notational convenience, we define \( \mathbf{H} \) as the channel matrix whose \((n, m)\)-th element is $H[n,m]$, and similarly for \( \mathbf{Y} \), \( \mathbf{X} \), and \( \mathbf{Z} \). Accordingly, the signal model \eqref{eq:Channel model} is equivalently expressed as
\begin{align}
    \mathbf{Y} = \mathbf{H} \circ \mathbf{X} + \mathbf{Z}.
\end{align}

5G NR specifies the configuration of OFDM slots.
For instance, in a standard OFDM resource block configuration with a normal CP and subcarrier spacing of $30$~kHz, each resource block comprises $12$ subcarriers in frequency and $14$ OFDM symbols in time.
With a typical Type-A DM-RS configuration for the PDSCH, the DM-RS occupies two OFDM symbols. These are commonly placed at symbol indices $2$ and $11$, depending on the configured pattern.
All resource elements in other symbols are typically allocated for data.
While we follow the 5G NR PDSCH configuration, the proposed method applies to arbitrary pilot patterns beyond this standard setting.

The OFDM channel estimation problem aims to estimate the channel matrix \( \mathbf{H} \), denoted by \( \hat{\mathbf{H}} \), given the observations \( {\bf{Y}} \) and the known pilot signal \( X[n, m] \) at \( (n, m) \in \CMcal{P} \).
The NMSE is commonly adopted as the performance metric, defined as
\begin{align} \label{eq: nmse def}
    \text{NMSE}(\hat{\mathbf{H}})= \frac{\mathbb{E}\left[ \|\mathbf{H}-\hat{\mathbf{H}} \|^{2}_{F} \right]}{\mathbb{E}\left[\| \mathbf{H}\|_{F}^{2}\right]}.
\end{align}
The estimation process involves denoising the pilot observations and interpolating the channel at the remaining resource elements in $\CMcal{D}$.

\section{Conventional Channel Estimation Approaches} \label{sec:conven}
In this section, we introduce conventional approaches to the OFDM channel estimation.
Existing methods can be broadly categorized into SP-based approaches and DNN-based approaches.
\subsection{SP-based Approach}

SP-based OFDM channel estimation strategies have been extensively studied in \cite{liu:survey:14}.
This work focuses on pilot-assisted estimation, which is widely adopted for its robustness and practicality.
The two most representative methods are the LS and LMMSE methods.

{\textbf{LS}}: The LS method minimizes the squared error between the received signal and the transmitted pilot signals, yielding the following closed-form solution at pilot positions:
\begin{align}
    \hat{H}[n,m] = \frac{Y[n,m]}{X[n,m]}, \quad (n,m) \in \CMcal{P}.
\end{align}
Channel coefficients at non-pilot positions $(n,m) \in \CMcal{T} \setminus \CMcal{P}$ are subsequently recovered via interpolation~\cite{coleri:tbc:02, edfors:tcom:98}.
While LS requires no prior statistical knowledge of the channel or noise, its performance degrades at low SNR due to the lack of any noise regularization.

{\textbf{LMMSE}}:
The LMMSE estimation uses prior channel statistics to minimize the expected MSE between the true channel and its estimate.
It corresponds to the Bayesian conditional mean of the channel given the pilot observations. This yields a statistically optimal solution to the channel estimation problem.
This is given by
\begin{align}
    \hat{\mathbf{H}}_{\text{LMMSE}}
    &= \mathop{\arg \min}_{\hat{\mathbf{H}}} \mathbb{E}\left[ \|\mathbf{H} - \hat{\mathbf{H}}\|_{ F}^2 \mid Y[n,m], (n,m) \in \CMcal{P} \right] \\
    &= \mathbb{E}\left[\mathbf{H} \mid Y[n,m], (n,m) \in \CMcal{P}\right], \label{eq:cond_mean}
\end{align}
where the expectation is taken over the randomness of ${\bf{H}}$ and ${\bf{Z}}$.
Assuming that \(\mathbf{H}\) and ${\bf{Z}}$ are jointly Gaussian, the conditional mean \eqref{eq:cond_mean} is given by the following linear expression:
\begin{align} \label{eq:mmse}
    \operatorname{vec}(\hat{\mathbf{H}}_{\text{LMMSE}}) &= \mathbb{E}\left[\mathbf{H}\mid Y[n,m], (n,m)\in\CMcal{P}\right] \nonumber \\
    &= \mathbf{R}_{{\mathbf{h}} {\bf{h}}_p}\mathbf{R}_{{\bf{y}}_p {\bf{y}}_p}^{-1}{{\bf{y}}_p},
\end{align}
where \(\mathbf{h} = \operatorname{vec}(\mathbf{H}) \in \mathbb{C}^{N M}\) denotes the vectorized full channel matrix, \({\bf{h}}_p \in \mathbb{C}^{L}\) represents the vectorized frequency response at the pilot positions, and $\mathbf{y}_p \in \mathbb{C}^L$ collects the received pilot signals $Y[n, m], \, (n,m) \in\CMcal{P}$.
\(\mathbf{R}_{\mathbf{h} {\bf{h}}_p} \in \mathbb{C}^{NM\times L}\) represents the cross-covariance matrix between \(\mathbf{h}\) and \({\bf{h}}_p \), while \(\mathbf{R}_{{\bf{y}}_p {\bf{y}}_p} \in \mathbb{C}^{L \times L}\) represents the auto-covariance matrix of \({\bf{y}}_p\).
More concretely, when ${\bf{Z}}$ is independent and identically distributed Gaussian with variance $\sigma^2$, \eqref{eq:mmse} is further derived as
\begin{align} \label{eq_lmmse_structure}
    \operatorname{vec}(\hat{\mathbf{H}}_{\text{LMMSE}}) &= \underbrace{\mathbf{R}_{\mathbf{h}{\bf{h}}_p} \left( \mathbf{R}_{{\bf{h}}_p {\bf{h}}_p} + \sigma^2 \mathbf{I} \right)^{-1}}_{\mathbf{W}_{\text{LMMSE}} \in \mathbb{C}^{NM \times L}} {\bf{y}}_p,
\end{align}
where ${\bf{W}}_{\text{LMMSE}}$ indicates the LMMSE filter.
Accordingly, the estimate corresponding to the specific resource element $(n,m)$ is given by
$\hat{H}\left[ n, m \right] = {\mathbf{R}_{H\left[n, m \right]{\bf{h}}_p} \left( \mathbf{R}_{{\bf{h}}_p {\bf{h}}_p} + \sigma^2 \mathbf{I} \right)^{-1}}{\bf{y}}_p$.
In this paper, the LMMSE refers to the estimator obtained in \eqref{eq:mmse}.

Under the joint Gaussian assumption, the LMMSE estimator is well known to achieve the optimal MSE. Unlike the LS estimator, it exploits knowledge of the covariance matrices $\mathbf{R}_{\mathbf{h}\mathbf{h}_p}$ and $ \mathbf{R}_{{\bf{h}}_p {\bf{h}}_p}$, which enable effective denoising and interpolation by accounting for channel correlations across time and frequency.

Due to the high dimensionality of the covariance matrix \(\mathbf{R}_{\mathbf{h} {\bf{h}}_p} \in \mathbb{C}^{NM\times L}\), a low-complexity approximation termed the 1D frequency-domain LMMSE (1D-LMMSE) was proposed, which neglects temporal correlation and operates solely in the frequency domain:
\begin{align}
    {\hat{\mathbf{H}}_{\text{1D-LMMSE}}}=\mathbf{R}_{\mathbf{H}[\CMcal{K}]\mathbf{H}[\CMcal{P}]}\left( \mathbf{R}_{\mathbf{H}[\CMcal{P}]\mathbf{H}[\CMcal{P}]}+\sigma^2\mathbf{I} \right)^{-1}\hat{\mathbf{H}}_{\text{LS}}, \label{1D FD-MMSE}
\end{align}
where \( \mathbf{H}[\CMcal{P}] \) denotes the frequency response at the pilot indices and \( \mathbf{H}[\CMcal{K}] \in \mathbb{C}^{N} \) denotes the full frequency response of all the subcarriers including the pilot component corresponding to \( \mathbf{H}[\CMcal{P}] \).

\begin{remark} \label{rem_sp} \normalfont
    Despite their theoretical foundations, SP-based methods face key practical limitations. The LMMSE requires accurate channel statistics that are difficult to obtain in practice, and its performance degrades under non-stationary conditions where the estimated covariance drifts from the true one, as quantified explicitly in Section~\ref{sec: nmse compar}.
\end{remark}

\subsection{DNN-based Approach}

To address the limitations of SP-based approaches, DNN-based approaches have been actively explored as a data-driven alternative.
The key idea is to leverage DNNs as a denoiser and also an interpolator for channel estimation. The DNN input is the noisy received pilot signal $\mathbf{y}_p$ and the output is the corresponding channel estimate $\hat{\mathbf{H}}$.
The DNN is trained to learn a nonlinear mapping between $\mathbf{y}_p$ and $\hat{\mathbf{H}}$.
Existing research on DNN-based approaches can be categorized according to the specific neural network architectures employed for channel estimation.

{\textbf{ChannelNet}}: The ChannelNet~\cite{soltani:commlett:19} is one of the initial DNN-based frameworks for OFDM channel estimation, treating the channel response as a 2D image.
A bilinear interpolation of the sparse LS estimates at pilot positions yields an initial low-resolution channel map.
This is then fed into a cascaded architecture.
A super-resolution convolutional neural network (i.e., SRCNN) recovers high-resolution spatial features. A denoising convolutional neural network (i.e., DnCNN) then applies residual learning and batch normalization to suppress residual noise and refine the estimate.

{\textbf{Channelformer}}:
The Channelformer \cite{luan2023channelformer} exploits the Attention Transformer architecture for OFDM channel estimation.
Unlike ChannelNet, Channelformer flattens the LS channel estimate for all pilot subcarriers into a single vector and feeds its real and imaginary parts as two separate input channels.
The real and imaginary parts of LS pilot estimates are concatenated into a sequence \(\mathbf{X} = \left[ \mathbf{e}_1^{\sf{T}}, \ldots, \mathbf{e}_L^{\sf{T}} \right] \in \mathbb{R}^{L \times 2}\), where $\mathbf{e}_i = \left[ \Re(p_i), \Im(p_i)\right] \in \mathbb{R}^{2}$, which serves as the encoder input.
The encoder applies the multi-head attention (MHA) mechanism~\cite{vaswani2017attention} to capture global dependencies across sparse pilot positions:

\begin{align} \label{eq:MHA}
    \text{head}_i &= \text{Attention}(\mathbf{Q}^{(i)}, \mathbf{K}^{(i)}, \mathbf{V}^{(i)}), \quad i = 1, \dots, h,\\
    \mathbf{Z} &= \text{Concat}(\text{head}_1, \dots, \text{head}_h)\mathbf{W}_O,
\end{align}
where \(\mathbf{Q}^{(i)}=\mathbf{Z}_{\text{in}}\mathbf{W}^{(i)}_Q, \ \mathbf{K}^{(i)}=\mathbf{Z}_{\text{in}}\mathbf{W}^{(i)}_K,\) and \(\mathbf{V}^{(i)}=\mathbf{Z}_{\text{in}}\mathbf{W}^{(i)}_V\) are obtained via learned linear projections, and the Scaled Dot-Product Attention is defined as
\begin{align} \label{eq:Attention}
    \text{Attention}(\mathbf{Q}^{(i)}, \mathbf{K}^{(i)}, \mathbf{V}^{(i)}) &= \operatorname{softmax}\left(\frac{\mathbf{Q}^{(i)}{\mathbf{K}^{(i)}}^{\sf{T}}}{\sqrt{d_k}}\right)\mathbf{V}^{(i)},
\end{align}
with $d_k$ denoting the number of pilot elements per symbol.
Channelformer replaces the standard position-wise feed-forward network (FFN) with a convolutional layer to inject local inductive bias alongside global Attention-based modeling:

\begin{align}
    \mathbf{Z}' = \text{Conv}(\mathbf{Z}), \quad \mathbf{H}_{\text{enc}} = \text{Norm}(\mathbf{Z} + \mathbf{Z}').
\end{align}
This design enables the encoder to simultaneously capture both global pilot dependencies and local channel variations.
The decoder maps the encoded representations to the full channel estimate through a residual convolutional neural network ($\CMcal{F}_{\text{ResCNN}}(\cdot)$):
\begin{align}
\hat{\mathbf{H}} &= \CMcal{F}_{\text{ResCNN}}(\mathbf{H}_{\text{enc}}| \theta_{\text{dec}}), \quad \hat{\mathbf{H}} \in \mathbb{C}^{N \times M},
\end{align}
where \( \theta_{\text{dec}} \) denotes the learnable parameters of the decoder.

\begin{remark} \label{rem_DNN} \normalfont
    DNN-based approaches offer two practical advantages over SP-based methods.
    They eliminate the need for explicit channel statistics and directly learn correlation structures from training data, enabling robust estimation even in high-mobility environments.
Nonetheless, these approaches rely on a large number of parameters and nonlinear activation functions at inference, and they provide no explicit mechanism to control computational overhead at deployment.

\end{remark}

\section{Attention-aided MMSE} \label{sec:ammse}

In this section, we propose the A-MMSE method for OFDM channel estimation.
Our key idea is to learn linear estimation filter coefficients using the Attention Transformer rather than directly employing a DNN to produce channel estimates.
Once trained, the learned filters are applied to noisy pilot signals to yield channel estimates through a linear operation, resembling the structure of the classical SP-based MMSE estimation approach.
In the following, we present a detailed explanation of the proposed A-MMSE method.
For clarity of exposition, we first describe the A-MMSE estimation mechanism, followed by the A-MMSE learning mechanism.

\subsection{A-MMSE Estimation Mechanism}

In the proposed A-MMSE approach, channel estimation follows conventional linear processing. Specifically, the estimated channel matrix $\hat{\mathbf{H}}_{\text{A-MMSE}}$ is obtained by
\begin{align} \label{eq:ammse}
    \operatorname{vec}(\hat{\mathbf{H}}_{\text{A-MMSE}}) &= {\mathbf{W}}_{\text{A-MMSE}} {\bf{y}}_p,
\end{align}
where ${\bf{W}}_{\text{A-MMSE}}$ represents the A-MMSE linear filter.
The A-MMSE filter ${\bf{W}}_{\text{A-MMSE}}$ is learned through our Attention Transformer architecture as follows:
\begin{align} \label{eq:learning_ammse}
{\bf{W}}_{\text{A-MMSE}} = \CMcal{F}_{\text{A-MMSE}}\left( \left[ \Re(Y[n, m]); \Im(Y[n,m]) \right],\right. \nonumber \\ \left.(n, m)\in \CMcal{P} \mid \theta_{\text{A-MMSE}} \right),
\end{align}
where $\CMcal{F}_{{\text{A-MMSE}}}(\cdot)$ represents the Attention Transformer network and \(\theta_{\text{A-MMSE}}\) denotes learnable parameters of A-MMSE.

\begin{remark} \label{rem_AMMSE} \normalfont
It is important to highlight several key points regarding the proposed A-MMSE method.
First, by leveraging the Attention Transformer to learn ${\bf{W}}_{\text{A-MMSE}}$, A-MMSE inherits the primary advantages of DNN-based approaches: powerful learning capability and no requirement for prior channel statistics.
We note that A-MMSE is positioned against practical plug-in LMMSE, in which the covariances $(\mathbf{R}_{\mathbf{h}\mathbf{h}_p}, \mathbf{R}_{\mathbf{h}_p\mathbf{h}_p})$
are estimated from finite, possibly non-stationary data; the oracle LMMSE with true population covariances remains optimal within the fixed linear class and is not claimed to be surpassed.
Section~\ref{sec:sim_set} additionally reports a per-frame oracle, computed from the true channel of each test frame, as a reference.
Second, once ${\bf{W}}_{\text{A-MMSE}}$ is learned, we only use linear computation as presented in \eqref{eq:ammse}, avoiding the computational complexity associated with nonlinear activation functions in the channel estimation process.
In this sense, the proposed A-MMSE is a hybrid approach that effectively combines SP-based and DNN-based approaches.
Further, in the A-MMSE framework, it is possible to adjust the number of effective parameters within ${\bf{W}}_{\text{A-MMSE}}$, enabling the A-MMSE to adapt its computational overhead depending on various environments. We further explore this in the next section.
\end{remark}

\begin{figure*}[t]
    \centering
    \includegraphics[width=1\linewidth]{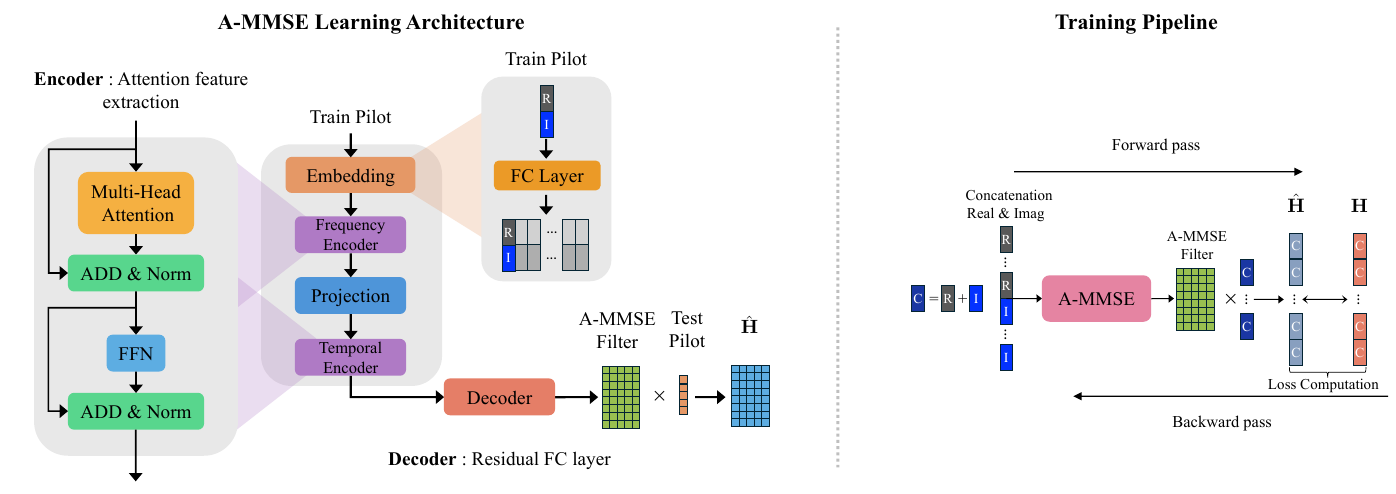}
    \caption{Architecture and end-to-end training pipeline of the proposed A-MMSE network for learning the MMSE filter.}
    \label{fig:Struc+Train}
\end{figure*}

\subsection{A-MMSE Learning Mechanism} \label{sub_learning Mechanism}

\subsubsection{Insights on Two-Stage Attention Encoder}

We now provide a detailed explanation of the A-MMSE learning mechanism $\CMcal{F}_{\text{A-MMSE}}(\cdot)$ \eqref{eq:learning_ammse}, whose overall architecture and training pipeline are depicted in Fig.~\ref{fig:Struc+Train}. Structurally inspired by recent encoder-decoder architectures based on the Attention Transformer, the proposed A-MMSE method incorporates theoretical insights derived from classical SP-based channel estimation approaches.
Specifically, as shown in the MMSE filter structure \eqref{eq_lmmse_structure}, the optimal linear filter operates by first whitening the pilot observations and then interpolating the full OFDM channel using the underlying frequency-temporal-domain channel correlation.
This highlights the importance of explicitly incorporating the channel's second-order statistics.
The proposed A-MMSE learning mechanism effectively captures the correlation structure inherent to OFDM channels.
Its detailed formulation is presented as follows.

Under the assumption of a wide-sense stationary uncorrelated scattering (WSSUS) environment and a separable scattering function \cite{Yoojin20152DLMMSE}, the second-order statistics of OFDM channels exhibit a separable correlation function across frequency and temporal domains. Formally, the channel correlation function between two channel taps separated by frequency index difference $\Delta n$ and time index difference $\Delta m$ is factorized as follows:
\begin{align}
\mathbf{R}_h(\Delta n, \Delta m) &= \mathbb{E}[\mathbf{H}[n, m] \mathbf{H}^*[n + \Delta n, m + \Delta m]] \nonumber \\
&= r_f(\Delta n) \cdot r_t(\Delta m), \label{eq_f_t_corr}
\end{align}
where \( r_f(\Delta n) \) and \( r_t(\Delta m) \) represent the frequency-domain and temporal-domain correlation, respectively.
Upon discretizing this correlation \eqref{eq_f_t_corr} over the OFDM grid,
the covariance matrix of the OFDM channel is given by
\begin{align}
    \mathbf{R}_{\text{full}} = \mathbb{E}[\mathbf{h} {\mathbf{h}}^{\mathsf H}] \approx \mathbf{R}_f \otimes \mathbf{R}_t \in \mathbb{C}^{NM\times NM}, \label{eq_corr}
\end{align}
where \(\mathbf{h}\) denotes \(\operatorname{vec}(\mathbf{H}) \in \mathbb{C}^{NM}\), \( \mathbf{R}_f \in \mathbb{C}^{N \times N} \) and \( \mathbf{R}_t \in \mathbb{C}^{M \times M} \) represent the frequency and temporal domain correlation matrices, respectively, as visualized in Fig.~\ref{fig:Encoder detail}.
We note that the approximation ($\approx$) in \eqref{eq_corr} arises because the WSSUS assumption may not perfectly hold in practical scenarios.
Importantly, this Kronecker structure serves as an inductive bias for encoder design rather than a hard constraint, allowing $\mathbf{W}_{\text{A-MMSE}}$ to adapt beyond this prior through end-to-end optimization.
The representation of the full covariance matrix ${\bf{R}}_{\text{full}}$ as the Kronecker product (\( \otimes \)) between the frequency-domain correlation ${\bf{R}}_{f}$ and the temporal-domain correlation ${\bf{R}}_t$ \eqref{eq_corr} implies that the frequency- and temporal-domain correlation structure is separable.
This allows for a compact and computationally efficient representation of the channel covariance matrix.
\begin{remark} \normalfont
    {The approximation in~\eqref{eq_corr} raises a natural concern: what happens when $\mathbf{R}_\mathrm{full}$ deviates from perfect Kronecker separability? By the Van Loan--Pitsianis decomposition~\cite{vanloan1993approximation}, any $\mathbf{R}_\mathrm{full}$ admits $\mathbf{R}_{\mathrm{full}} = \sum_{k=1}^{K} \sigma_k\, \mathbf{R}_f^{(k)} \otimes \mathbf{R}_t^{(k)}$. The single-term model in~\eqref{eq_corr} retains the dominant component  $\sigma_1\, \mathbf{R}_f^{(1)} \otimes \mathbf{R}_t^{(1)}$, which the two-stage encoder captures, while the residual $\sum_{k \ge 2}\sigma_k \mathbf{R}_f^{(k)}\otimes \mathbf{R}_t^{(k)}$ is absorbed through end-to-end optimization of $\mathbf{W}_{\text{A-MMSE}}$, which carries no separability constraint. The encoder thus exploits the dominant Kronecker structure, and the learned filter handles the rest.
    }
\end{remark}
Building on this insight, we propose the following two-stage encoder design.

\begin{itemize}
    \item \textbf{Frequency Encoder:} In the Frequency Encoder stage, the pilot observations are processed along the subcarrier domain, extracting features on the frequency-domain correlation structure at pilot indices, corresponding to the rows of the matrix \( \mathbf{R}_f\left[\CMcal{P},: \right] \).

    \item \textbf{Temporal Encoder:} In the Temporal Encoder stage, the output of the Frequency Encoder is subsequently processed along the OFDM symbol domain, capturing temporal-domain correlation structures at pilot indices, corresponding to the rows of the matrix \( \mathbf{R}_t \left[\CMcal{P},: \right] \).
\end{itemize}

This two-stage encoder enables the model to learn the joint frequency--temporal correlation structure; each stage uses a single encoder block, which is critical for accurate MMSE estimation.
Crucially, we note that this encoder design represents a key distinction of the proposed A-MMSE learning mechanism compared to conventional DNN-based methods.
Specifically, existing approaches typically adopt standard learning architectures such as CNNs \cite{soltani:commlett:19} or Attention Transformers \cite{luan2023channelformer}, without explicitly tailoring them to the statistical properties inherent in OFDM channels.
In contrast, our method explicitly incorporates the inherent temporal and frequency-domain correlation structures of OFDM channels into the learning process.
This integration not only enhances learning efficiency but also improves the model's capability to generalize under practical channel conditions.
In this sense, the proposed A-MMSE approach is distinguished not only by its design, which enables the learning of a linear estimation filter rather than directly estimating the OFDM channel matrix (as illustrated in Fig.~\ref{fig:comparison}), but also by its advanced learning mechanism that explicitly reflects the inherent correlation structure of OFDM channels.

\subsubsection{Multi-Head Self-Attention Mechanism}

Building upon the proposed two-stage encoder structure,
we use the MHA mechanism in \eqref{eq:MHA}--\eqref{eq:Attention} to extract encoded features.
In particular, the Frequency Encoder captures correlations between pilot observations and individual subcarriers, thus its number of attention heads is fixed to the pilot count within a single OFDM symbol.
The Temporal Encoder then extends the resulting frequency-domain features along the symbol axis to model the inter-symbol correlation. Accordingly, the number of attention heads is set equal to the total number of OFDM symbols considered.
We note that a multi-head self-attention layer with $N_h$ heads has the capacity to express convolutional operations whose receptive field is controlled by the number of heads~\cite{rel:conv:att:2020}, so that the head budgets $L$ and $M$ allow our encoders to represent both local (cluster-level) and global correlation patterns within a single attention stage.
The statistical features extracted by the two-stage encoder are subsequently passed through a residual fully connected decoder, which maps them into the coefficient space of the A-MMSE filter.
The detailed process is as follows.

\begin{figure}[]
    \centering
    \includegraphics[width=1\linewidth]{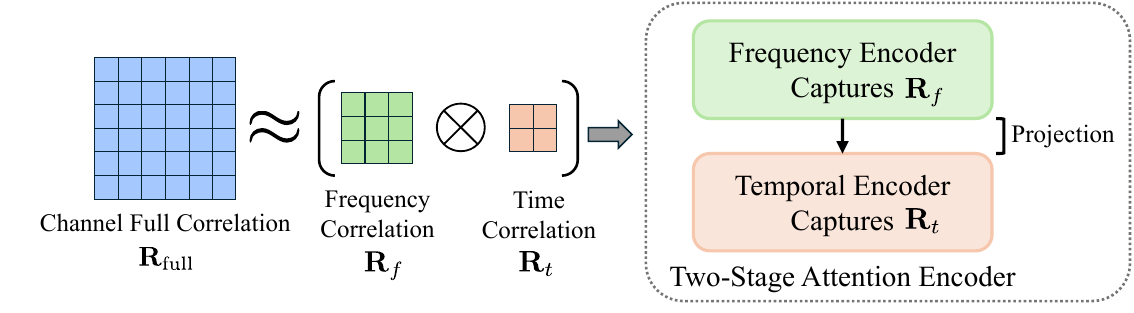}
    \caption{Detailed structure of the proposed two-stage Attention encoder.}
    \label{fig:Encoder detail}
\end{figure}

\textbf{Frequency Encoder}:
The detailed structure of the two-stage encoder is depicted in Fig.~\ref{fig:Encoder detail}. In the first stage, a linear embedding layer transforms the raw pilot input data into a high-dimensional embedding space, where the embedding dimension matches the number of subcarriers. This embedding is essential to facilitate the subsequent MHA mechanism, which effectively captures the frequency-domain dependencies inherent among the pilot signals.
Specifically, the noisy pilot vector is given by \(\mathbf{x}_{\text{in}}=\left[x_{\text{1}}, x_{\text{2}}, \dots, x_{2L} \right]^{\mathsf{T}} \in \mathbb{R}^{2L}\). Here, each scalar \(x_i\) is linearly embedded into a high-dimensional vector as: \(\mathbf{x}_{\text{emb}, i} = x_i \cdot \mathbf{P}_{\text{emb}} + \mathbf{b}_{\text{emb}}\), where \(\mathbf{P}_{\text{emb}}\in \mathbb{R}^{1 \times d_e}\) is a learnable projection matrix and  \(\mathbf{b}_{\text{emb}}\in\mathbb{R}^{d_e}\) is a learnable bias vector. Stacking all \(\mathbf{x}_{\text{emb}, i}\), we have
\begin{align}
    \mathbf{X}_{\text{Emb}} &= [ \mathbf{x}_{\text{emb,1}}^{\mathsf{T}}, \mathbf{x}_{\text{emb, 2}}^{\mathsf{T}}, \cdots, \mathbf{x}_{\text{emb}, 2L}^{\mathsf{T}}]^{\mathsf{T}} \in \mathbb{R}^{2L \times d_e}.
\end{align}
The embedding dimension \(d_e\) is set equal to $N$ (the number of subcarriers), ensuring that the resulting embedded representations effectively capture complex frequency-domain characteristics.

Subsequently, the embedding matrix $\mathbf{X}_{\text{Emb}}$ is fed into the MHA module \(\left(\CMcal{F}_{\text{MHA}}(\cdot) \right)\) with the number of heads fixed at $L/2$, where $L/2$ corresponds to the number of pilots within a single DM-RS symbol (one head per pilot).
Each complete pilot is split into real and imaginary parts, resulting in the $2L$ column dimension of $\mathbf{X}_{\mathrm{Emb}}$.
For example, with Type-A DM-RS and additional position $=1$ (i.e., two DM-RS OFDM symbols) using a comb-2 pattern over 6 resource blocks (RBs), which comprise 72 subcarriers, there are $72/2=36$ pilots per DM-RS symbol -- equal to the number of Frequency Encoder heads -- and hence $L = 36 \times 2 = 72$ pilots in total. In general, the total pilot count is
\begin{align}
    L=\left( \frac{N_{\mathrm{RB}}\cdot N_{\mathrm{SC/RB}}}{2}  \right) \times N_{\text{DM-RS sym}},
\end{align}
where $N_{\mathrm{RB}}$ is the number of resource blocks, $N_{\mathrm{SC/RB}}=12$ is the number of subcarriers per RB, and $N_{\text{DM-RS sym}}$ denotes the number of OFDM symbols carrying the DM-RS.
The per-symbol pilot count $(N_{\mathrm{RB}} \cdot N_{\mathrm{SC/RB}})/2$ sets the number of Frequency Encoder heads.
The operation of the MHA module $\CMcal{F}_{\text{MHA}}(\cdot)$ is described in \eqref{eq:MHA}--\eqref{eq:Attention}.
We note that this allows each head to learn distinct frequency-domain correlation patterns.
Following the MHA module, Add \& Layer Normalization is applied, followed by an FFN, and subsequently another Add \& Layer Normalization, in sequential order. Details on Add \& Norm and FFN are provided in \cite{vaswani2017attention}. After these processes, the encoder extracts frequency-domain statistical information, which can be formally represented as
\begin{align}
    \mathbf{Y}_{\text{Freq}} = \CMcal{E}_{\text{Frequency}}\left(\CMcal{F}_{\text{MHA}}(\mathbf{X}_{\text{Emb}}| \theta_{\text{MHA, Freq}})|  \theta_{\text{Freq}} \right),
\end{align}
where \(\mathbf{Y}_{\text{Freq}}\in \mathbb{R}^{2L\times N}\) represents features encoded by the Frequency Encoder, \(\CMcal{E}_{\text{Frequency}}\) represents the Frequency Encoder module, and $\theta_{\text{MHA, Freq}}$ and $\theta_{\text{Freq}}$ denote the learnable parameters of the MHA module and the combined parameters of the two Add \& Layer Normalization layers and the FFN, respectively.
Correspondingly, the overall operation \( \mathbf{x}_{\text{in}} \mapsto \mathbf{Y}_{\text{Freq}}\) yields a real-valued matrix, where each row corresponds to a pilot position and columns align with subcarrier-specific embedding dimensions, explicitly capturing both local and global dependencies in the frequency domain.

\textbf{Temporal Encoder}: In the second stage, a linear projection layer transforms the features encoded in the previous stage into a high-dimensional frequency and temporal domain, where the projection dimension is $NM$.
The encoded feature matrix is given as \(\mathbf{Y}_{\text{Freq}} = \left[\mathbf{y}_{\text{Freq}, 1};\mathbf{y}_{\text{Freq}, 2};\cdots;\mathbf{y}_{\text{Freq}, 2L} \right] \in \mathbb{R}^{2L\times N}\). Here, each row vector \(\mathbf{y}_{\text{Freq}, i}\) is linearly projected as \(\mathbf{x}_{\text{proj}, i}=\mathbf{y}_{\text{Freq}, i}\cdot \mathbf{P}_{\text{proj}}+\mathbf{b}_{\text{proj}}\), where \(\mathbf{P}_{\text{proj}}\in \mathbb{R}^{N \times d_p}\) is a learnable projection matrix and \(\mathbf{b}_{\text{proj}}\in \mathbb{R}^{d_p}\) is a learnable bias vector.
Stacking all projected vectors \(\mathbf{x}_{\text{proj}, i}\), this yields the matrix given by
\begin{align}
    \mathbf{X}_{\text{Proj}} &= [ \mathbf{x}_{\text{proj,1}}^{\mathsf{T}}, \mathbf{x}_{\text{proj, 2}}^{\mathsf{T}}, \cdots, \mathbf{x}_{\text{proj}, 2L}^{\mathsf{T}} ]^{\mathsf{T}}  \in \mathbb{R}^{2L \times d_p}.
\end{align}
We note that the projection dimension $d_p$ is set to $NM$, ensuring that the model captures the frequency-temporal correlation of the OFDM channel, as formulated in \eqref{eq_corr}.

Next, the projected matrix $\mathbf{X}_{\text{Proj}}$ is fed into another MHA module, with the number of heads fixed at $M$ (one head for each OFDM symbol).
Similar to the above, this enables each head to effectively learn distinct temporal correlation patterns.
The subsequent process is identical to that of the Frequency Encoder.

After completing all the steps, the frequency-temporal statistical information extracted by the Temporal Encoder can be represented as follows:
\begin{align}
    \mathbf{Y}_{\text{Temp}} = \CMcal{E}_{\text{Temporal}}\left(\CMcal{F}_{\text{MHA}}(\mathbf{X}_{\text{Proj}}| \theta_{\text{MHA, Temp}})|  \theta_{\text{Temp}} \right),
\end{align}

where \(\mathbf{Y}_{\text{Temp}}\in \mathbb{R}^{2L\times NM}\) represents features encoded by the Temporal Encoder, \(\CMcal{E}_{\text{Temporal}}\) represents the Temporal Encoder module, with \( \theta_{\text{MHA, Temp}} \) and \(\theta_{\text{Temp}}\) being the learnable parameters of the MHA module and the combined parameters of the two Add \& Layer Normalization layers and the FFN, respectively.
The overall operations \(\mathbf{Y}_{\text{Freq}} \mapsto \mathbf{Y}_{\text{Temp}}\) yield a real-valued matrix,  whose rows correspond to pilot positions and whose columns align with subcarrier-symbol-specific embedding dimensions. This explicitly captures both local and global dependencies in the frequency and temporal domain of the OFDM channel. The frequency-temporal features encoded by the two-stage encoder serve as inputs to the decoder.

\textbf{Residual Fully Connected Decoder}:
Now we describe the decoder structure.
The decoder of the A-MMSE model employs a residual fully connected (ResFC) module.
Through this, we aim to reconstruct the final A-MMSE filter from the encoded frequency-temporal features.

The decoder input, i.e., $\mathbf{Y}_{\text{Temp}}$, is processed through multiple fully connected layers with residual connections to ensure stable training and efficient transformation. The output dimension of the decoder remains consistent with its input dimension. This is presented as
\begin{align}
\mathbf{Y}_{\text{Dec}} = \CMcal{F}_{\text{ResFC}}\left( \mathbf{Y}_{\text{Temp}}|\theta_{\text{ResFC}} \right) \in \mathbb{R}^{2L \times NM},
\end{align}
where \(\mathbf{Y}_{\text{Dec}}\in \mathbb{R}^{2L \times NM}\) represents the A-MMSE filter coefficients mapped by the ResFC decoder, and \(\theta_{\text{ResFC}}\) denotes the learnable decoder parameters.

Once $\mathbf{Y}_{\text{Dec}}$ is obtained, it is partitioned into the real and imaginary components as follows:
\begin{align}
{\mathbf{W}}_{\text{real}} &= \mathbf{Y}_{\text{Dec}}{[1:L,:]} \in \mathbb{R}^{L \times N M}, \\
{\mathbf{W}}_{\text{imag}} &= \mathbf{Y}_{\text{Dec}}{[L+1:2L,:]} \in \mathbb{R}^{L \times N M}.
\end{align}
Then, the final complex-valued filter matrix is reconstructed by combining these two components as
\begin{align}
    {\mathbf{W}}_{\text{out}} &= {\mathbf{W}}_{\text{real}} + j\, {\mathbf{W}}_{\text{imag}}, \
    {\mathbf{W}}_{\text{A-MMSE}} = {\mathbf{W}}_{\text{out}}^{\sf{T}} \in \mathbb{C}^{N M \times L}.
\end{align}
This completes the end-to-end A-MMSE learning pipeline.

\subsubsection{Training Method} \label{sec: training method}

The proposed A-MMSE model is trained in an end-to-end fashion as shown in the training pipeline of Fig.~\ref{fig:Struc+Train}.
The model parameters are optimized by minimizing a certain loss function, which quantifies the discrepancy between the estimated and true channel responses across the entire dataset:
\begin{align} \label{eq:loss}
\text{Loss}= \frac{1}{ N_{\CMcal{B}} \times \CMcal{B}}\sum_{j=1}^{N_{\CMcal{B}}}\sum_{i=1}^{\CMcal{B}} \CMcal{L}\left(\hat{\mathbf{H}}_{i, j}, \mathbf{H}_{i, j}\right),
\end{align}
where $N_{\CMcal{B}}$ denotes the number of mini-batches and $\CMcal{B}$ denotes the mini-batch size.
In addition, $\mathbf{H}_{i,j}$ and $\hat {\bf{H}}_{i,j}$ represent the corresponding true channel matrix and estimated channel matrix, respectively.
For the loss function $\CMcal{L}(\cdot)$, MSE or Huber loss can be chosen; the MSE loss is used in the simulations of Section~\ref{sec:sim}.
The loss function \eqref{eq:loss} is minimized via backpropagation and the Adam optimizer \cite{kingma2014adam}.

After training, one final A-MMSE filter is produced from the entire training dataset. This A-MMSE filter remains fixed and is consistently applied to all test data without further modification.

\subsection{Discussions} \label{sec: ammse discuss}
Now we provide further insights regarding the proposed A-MMSE method.
The proposed A-MMSE is fundamentally distinguished from existing DNN-based channel estimation approaches.
To be specific, the proposed A-MMSE method learns a linear filter matrix \( \mathbf{W}_{\text{A-MMSE}} \) during a dedicated training phase.
As a result, once trained, channel estimation reduces to a single matrix-vector multiplication, as shown in \eqref{eq:ammse}.
This enables highly efficient inference with minimal computational cost, requiring only \( 2(NM \times L) \) real-valued parameters to store \( \mathbf{W}_{\text{A-MMSE}} \).
In contrast, as discussed earlier, conventional DNN-based methods \cite{luan2023channelformer, soltani:commlett:19} learn a neural network that maps pilot inputs directly to channel outputs.
Accordingly, each inference (i.e., channel estimation) involves propagating the input through multiple layers of nonlinear operations, leading to higher inference complexity.
Specifically, the A-MMSE performs channel estimation using only linear operations with approximately 70,000 \(\left(=NM \times L\right)\) parameters, whereas Channelformer requires about 110,000 parameters and includes nonlinear components, such as GELU activations and convolutional layers.
Since A-MMSE inference is a fixed linear map, the benchmark for the deployed filter is the population-optimal filter within the class of fixed linear estimators, not an unrestricted nonlinear MMSE estimator.
For stationary channels, this benchmark is the Wiener filter. Appendix~\ref{app:mse} characterizes this benchmark and the penalty of any fixed filter relative to it. In non-stationary settings, A-MMSE shares the limit of the fixed linear class with plug-in LMMSE: neither can go below the per-frame oracle of Section~\ref{sec: nmse compar}. Section~\ref{sec:sim} compares the mismatch penalties of the two on test frames that follow the training frames in time.

Nevertheless, storing the full matrix $\mathbf{W}_{\text{A-MMSE}}$ and performing large-scale matrix-vector multiplications can be burdensome for resource-constrained receivers as $N$, $M$, or $L$ increase.
It is therefore desirable to adapt the computation and memory requirements of A-MMSE to available hardware resources.
The following section introduces a rank-adaptive extension that achieves this by adjusting the filter rank at deployment time.

\begin{figure}
    \centering
    \includegraphics[width=1\linewidth]{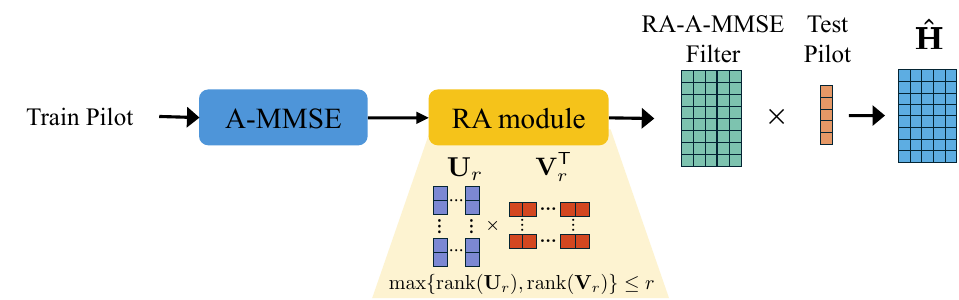}
    \caption{RA-A-MMSE architecture: a rank-adaptation (RA) module is appended to A-MMSE to flexibly adjust the rank of the learned A-MMSE filter. The A-MMSE is trained beforehand and held fixed; only the matrices of the RA module are learned, separately for each offered rank $r$.}
    \label{fig:RA-A-MMSE}
\end{figure}

\section{Rank-Adaptive A-MMSE} \label{sec:ra_ammse}

In this section, by extending the A-MMSE, we introduce the RA-A-MMSE framework, whose architecture is illustrated in Fig.~\ref{fig:RA-A-MMSE}.
Note that RA-A-MMSE is designed not to improve NMSE over A-MMSE, but to provide a flexible accuracy--complexity trade-off for resource-constrained UEs.
The RA-A-MMSE incorporates a rank-adaptation mechanism that adaptively reduces the filter matrix rank according to resource constraints, such as a predefined maximum rank \((r)\) or a limit on the number of filter parameters.
Importantly, the rank $r$ is a deployment-time parameter determined by the UE's hardware resource availability (e.g., memory and compute budget), not by instantaneous channel conditions such as SNR or Doppler shift; channel conditions vary on a per-slot basis, whereas hardware capability is fixed per device.
By flexibly adjusting the rank to satisfy these constraints, the RA-A-MMSE effectively reduces computational and memory overhead, allowing the filter to operate efficiently even under stringent computational resource limitations.
Despite operating at a reduced rank, the RA-A-MMSE still maintains high estimation accuracy.
It thus offers a practical and scalable solution for resource-constrained OFDM receivers, achieving a favorable balance between performance and hardware efficiency.

\begin{table*}[b]
\centering
\caption{Configuration Parameters for HSR and Semi-Urban Scenarios}
\label{tab:cost_config}
\renewcommand{\arraystretch}{1.0} %
\begin{tabular}{lcccccc}
\hline
\textbf{Scenario} & \textbf{Freq. ($f_c$)} & \textbf{Delay Spread} & \textbf{SCS} & \textbf{Velocity ($v$)} & \textbf{Max Doppler} & \textbf{Propagation} \\ \hline
\textbf{HSR} & 5 GHz & 100 ns & 60 kHz & 350 km/h & $\approx 1620.4$ Hz & Strong LoS ($K=13$ dB) \\
\textbf{Semi-Urban} & 3.5 GHz & 1000 ns & 30 kHz & 40 km/h & $\approx 129.6$ Hz & Rich Scattering ($K=3$ dB) \\ \hline
\end{tabular}
\end{table*}

To explain the key idea of RA-A-MMSE, we begin by examining how rank reduction alleviates computational complexity.
Recall that the A-MMSE filter \( \mathbf{W}_{\text{A-MMSE}} \in \mathbb{C}^{NM \times L} \) is a linear transformation from the pilot observations to the channel estimate.
Accordingly, without any rank-specific modification, the A-MMSE channel estimation requires linear operations proportional to \( 8NM \times L \), accounting for both real and imaginary components in complex-valued multiplications. Now suppose that the rank of \( \mathbf{W}_{\text{RA-A-MMSE}} \) is reduced to \( r \).
Then the following rank-$r$ factorization is allowed:
\begin{align}
    \mathbf{W}_{\text{RA-A-MMSE}} = \mathbf{A}\mathbf{B}^{\sf T}, \quad \mathbf{A} \in \mathbb{C}^{NM \times r},\ \mathbf{B} \in \mathbb{C}^{L \times r}.
\end{align}
In this case, the estimation process can be decomposed into two sequential steps:
\begin{align} \label{eq: ra est}
    \tilde{\bf{y}}_p = \mathbf{B}^{\sf T} \mathbf{y}_p, \; \operatorname{vec}(\hat{\mathbf{H}}_{\text{RA-A-MMSE}}) = \mathbf{A} \tilde{\bf{y}}_p,
\end{align}
where \( \mathbf{y}_p \in \mathbb{C}^L \) is the pilot input and $\operatorname{vec}(\hat{\mathbf{H}}_{\text{RA-A-MMSE}})$ is the channel estimate.
This decomposition reduces the computational complexity to approximately \( 8rL + 8NMr \) real-valued operations, which is significantly lower than the full-rank case when \( r \ll \min(NM, L) \).

To achieve the rank reduction in the RA-A-MMSE, we devise a rank-adaptive (RA) module. The RA module consists of one or more cascaded pairs of learnable matrices. We first describe a single pair, $\mathbf{U}_r \in \mathbb{R}^{L \times r}$ and $\mathbf{V}_r \in \mathbb{R}^{L \times r}$, where $r$ denotes the desired rank.
The RA module is appended to the A-MMSE obtained once by the procedure of Section~\ref{sec: training method} and does not require modifying or retraining the A-MMSE.
During training of the RA module, the encoder, the decoder, and the filter $\mathbf{W}_{\text{A-MMSE}}$ are held fixed. Only the matrices of the RA module are updated.
For each rank $r$ that the BS chooses to offer, the RA module for that rank is trained separately, offline at the BS. Its matrices are optimized to minimize the MSE of the channel estimate produced by the resulting filter $\mathbf{W}_{\text{RA-A-MMSE}}$.
The loss function~\eqref{eq:loss} and the training data are the same as those used for the A-MMSE. The retained subspace thus captures the most informative components of $\mathbf{W}_{\text{A-MMSE}}$.
With $\mathbf{U}_r$ and $\mathbf{V}_r$, the A-MMSE filter matrix ${\bf{W}}_{\text{A-MMSE}}$ is modified to  $\mathbf{W}_{\text{RA-A-MMSE}}
    = \mathbf{W}_{\text{A-MMSE}} \,{\mathbf{U}_r\mathbf{V}_r}^{\mathsf{T}}$.
Then the rank of $\mathbf{W}_{\text{RA-A-MMSE}}$ is bounded by
\begin{align} \label{eq:ra_rank}
    \begin{split}
        \operatorname{rank}\bigl(
            \mathbf{W}_{\text{A-MMSE}}\,
            \mathbf{U}_r\,
            \mathbf{V}_r^{\mathsf{T}}
        \bigr)
        \;\le\;
        \min\{
            \operatorname{rank}(\mathbf{W}_{\text{A-MMSE}}), \\
            \quad \operatorname{rank}(\mathbf{U}_r),\
            \operatorname{rank}(\mathbf{V}_r)
        \} \le r,
    \end{split}
\end{align}
where the last inequality holds if $r \le \min\{NM,L\}$.
As a result, the rank of $\mathbf{W}_{\text{RA-A-MMSE}} $ is bounded by $r$, allowing the effective filter rank to be adaptively controlled via the dimensions of \(\mathbf{U}_r\) and \(\mathbf{V}_r\).
This enables the RA-A-MMSE to flexibly adapt to available system resources.
The RA module may also consist of $S > 1$ cascaded pairs $\mathbf{U}_{r_s}, \mathbf{V}_{r_s} \in \mathbb{R}^{L \times r_s}$, $s = 1, \ldots, S$, with $r_1 \ge \cdots \ge r_S  = r$. In this case, the filter becomes
\begin{align} \label{eq:ra_cascade}
    \mathbf{W}_{\text{RA-A-MMSE}} = \mathbf{W}_{\text{A-MMSE}}\,\mathbf{U}_{r_1}\mathbf{V}_{r_1}^{\sf T}\cdots\mathbf{U}_{r_S}\mathbf{V}_{r_S}^{\sf T}.
\end{align}
Its rank is bounded by $r_S = r$ by the same argument, since the rank of a product does not exceed the rank of any factor.
The trained parameters amount to $2L\sum_{s=1}^{S} r_s$ real values, which reduces to $2Lr$ for a single pair. The number of cascaded pairs and the intermediate ranks $r_1, \ldots, r_{S-1}$ are configuration choices.
They do not change the rank-$r$ factorization or the deployment described below; the size of the deployed filter is determined by the final rank $r$ alone.
\begin{figure}[t]
    \centering
    \includegraphics[width=1\linewidth]{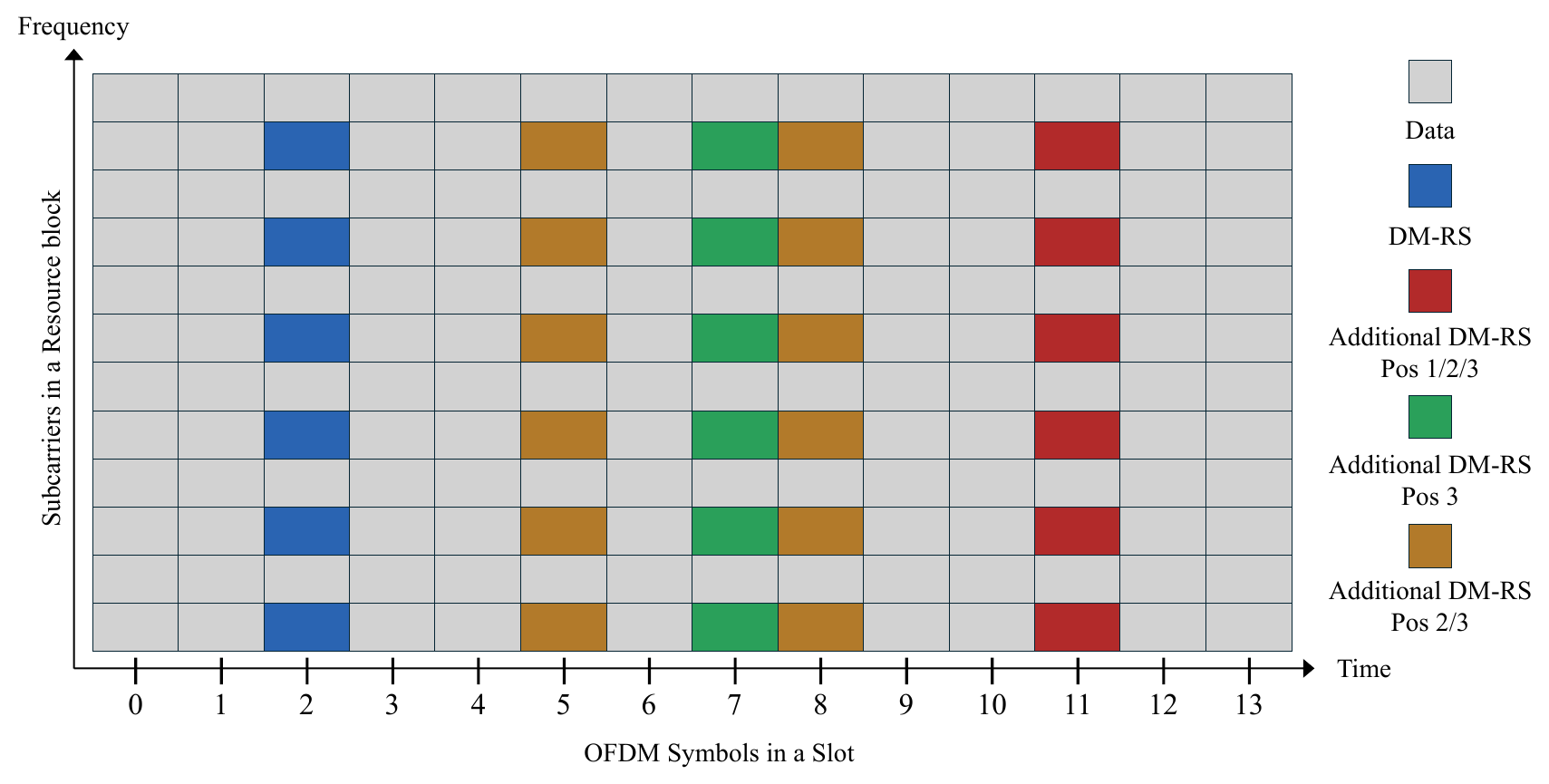}
    \caption{5G NR time-frequency resource block with Type-A DM-RS.
    }
    \label{fig:DM-RS}
\end{figure}
Compared to the A-MMSE, the computational cost reduction of the RA-A-MMSE can be quantified by the ratio:
\begin{align}
    \text{Complexity Cost Ratio:}\; \frac{\text{RA-A-MMSE}}{\text{A-MMSE}}\approx \frac{NMr+Lr}{NML}.
\end{align}

Considering an OFDM channel comprising $6$ RBs (minimum bandwidth assignment), i.e., \(N=72\), \(M=14\),
the RA-A-MMSE with $r = 12$ reduces the computational cost to approximately $18\%$ of the full-rank A-MMSE for $L=72$.
After the RA module has been trained for a rank $r$, the BS forms the filter $\mathbf{W}_{\text{RA-A-MMSE}}$ defined above, whose rank is at most $r$.
It then obtains $\mathbf{A} \in \mathbb{C}^{NM \times r}$ and $\mathbf{B} \in \mathbb{C}^{L \times r}$ by a rank-$r$ factorization of this matrix, so that $\mathbf{A}\mathbf{B}^{\sf T} = \mathbf{W}_{\text{RA-A-MMSE}}$. This factorization is computed once, offline, and does not involve training. The BS stores the full $\mathbf{W}_{\text{A-MMSE}}$ and one $(\mathbf{A}, \mathbf{B})$ pair per offered rank.
The set of offered ranks is a design choice of the BS; Section~\ref{sec:sim} evaluates three representative ranks.
In practice, the BS selects a single rank $r$ based on UE capability signaling, and deploys only the corresponding $(\mathbf{A}, \mathbf{B})$ pair---approximately $r/L$ of the full-rank filter. Neither training nor computation of $(\mathbf{A}, \mathbf{B})$ is required at deployment; the UE executes only the two sequential steps in~\eqref{eq: ra est}.
A UE whose capability supports the full rank receives $\mathbf{W}_{\text{A-MMSE}}$ itself, and the RA module is not used.
Offering a rank that has not been prepared before requires only one additional training of the RA module for that rank. The A-MMSE itself remains unchanged. Since the matrices of the RA module are learned to minimize the estimation error, this substantial complexity reduction is achieved with minimal channel estimation performance loss.
We demonstrate this in the next section for the $50\%$, $25\%$, and $10\%$ rank configurations ($r = 36, 18, 7$ for $L = 72$). Each of these configurations was obtained by a separate RA-module training with the A-MMSE fixed, using $S = 1$, $2$, and $3$ cascaded pairs, respectively.

\begin{figure*}[t]
    \centering
    \subfigure[NMSE performance under the Semi-Urban (SU) scenario.]{
        \includegraphics[width=0.48\textwidth]{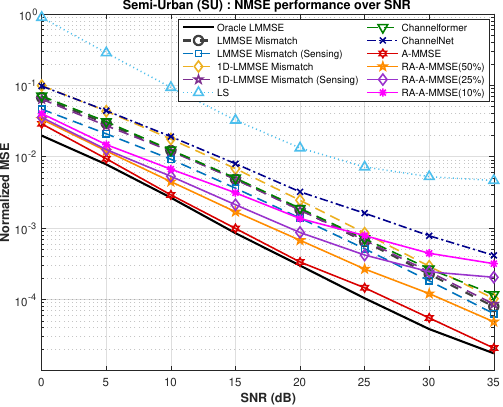}
        \label{fig:NMSE_SU}}
    \hfill
    \subfigure[NMSE performance under the High-Speed Rail (HSR) scenario.]{
        \includegraphics[width=0.48\textwidth]{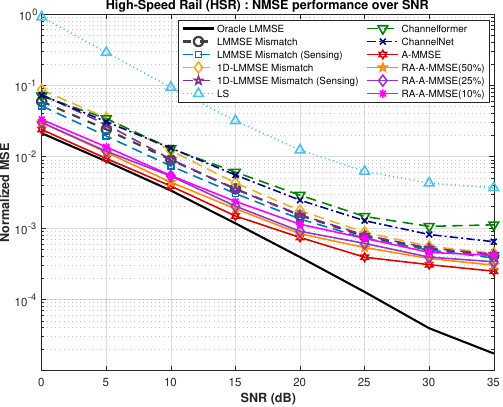}
        \label{fig:NMSE_HSR}}
    \caption{NMSE performance comparison of various channel estimation methods over SNR under COST2100 channel models.}
    \label{fig:NMSE_COST}
\end{figure*}

\begin{remark} \label{sec:RA_remark} \normalfont
    The proposed A-MMSE exploits sparsity in two ways. First, the MHA mechanism induces soft sparsity by assigning higher attention weights to dominant features, consistent with the separable covariance structure ($\mathbf{R}_{\mathrm{full}} \approx \mathbf{R}_f \otimes \mathbf{R}_t$) of wideband wireless channels. Second, RA-A-MMSE learns a rank-constrained modification of the fixed A-MMSE filter through the RA module, with rank at most $r$ as shown in \eqref{eq:ra_rank}--\eqref{eq:ra_cascade}. The resulting accuracy--complexity trade-off is evaluated in Section~\ref{sec:sim}.
\end{remark}

\section{Simulation Results} \label{sec:sim}

\subsection{Simulation Setup} \label{sec:sim_set}

Numerical simulations follow the 5G NR DM-RS configuration specified in 3GPP TS 38.211 \cite{3GPP-TS-38.211}. As illustrated in Fig.~\ref{fig:DM-RS}, the DM-RS time-frequency density is adapted to channel dynamics, balancing estimation accuracy against pilot overhead. We evaluate channel estimation performance using the geometry-based COST2100 model \cite{COST2100} in Semi-Urban (SU) and High-Speed Rail (HSR) scenarios (Table~\ref{tab:cost_config}).
The COST2100 model employs clustered multipath components with frequency and temporal correlations that are not, in general, Kronecker-separable. This makes it a demanding testbed for assessing robustness beyond the idealized WSSUS and separability assumptions.

For each scenario, $44{,}000$ continuous OFDM frames were generated with dynamic Doppler updates, creating a non-stationary environment where the global WSSUS assumption holds only over short local intervals.
The dataset is split temporally: the first $40{,}000$ frames serve for training ($36{,}000$) and validation ($4{,}000$), and the subsequent $4{,}000$ frames form the test set.
The model was implemented in TensorFlow on a single NVIDIA RTX 4090 GPU. All source code is available at \cite{A-MMSE-GitHub}.
The subcarrier spacings in Table~\ref{tab:cost_config} were chosen to evaluate the trade-off between Doppler resolution and robustness to channel time-variation.
Pilot symbols are set to one, so that $\mathbf{y}_p = \mathbf{h}_p + \mathbf{z}_p$ with $\mathbb{E}[\mathbf{z}_p] = \mathbf{0}$ and $\mathbb{E}[\mathbf{z}_p\mathbf{z}_p^{\mathsf H}] = \sigma^2\mathbf{I}$.
Baselines include SP-based estimators (LS, LMMSE mismatch, 1D-LMMSE, and sensing-aided estimation~\cite{Qing2024IoTJ}) and DNN-based models (ChannelNet~\cite{soltani:commlett:19} and Channelformer~\cite{luan2023channelformer}).
The LMMSE mismatch filter inserts the sample second moments $\tilde{\mathbf{R}}_{\mathbf{h}\mathbf{h}_p}$ and $\tilde{\mathbf{R}}_{\mathbf{h}_p\mathbf{h}_p}$ of the true channels of the $36{,}000$ training frames into the LMMSE filter~\eqref{eq_lmmse_structure}. The resulting single filter is applied unchanged to all $4{,}000$ test frames; its mismatch with the true time-varying statistics is quantified in Section~\ref{sec: nmse compar}.
A-MMSE is trained with the MSE loss in~\eqref{eq:loss} on the $36{,}000$ frames.
The Oracle LMMSE is shown as a reference, not as a baseline. For every test frame (one OFDM slot, i.e., one $N \times M$ time-frequency grid), it is recomputed from the second-order moments of that frame's true channel.
It presupposes the channel it estimates and is therefore not realizable by a receiver; its role as a lower bound is established in Section~\ref{sec: nmse compar}. The noise variance $\sigma^2$ is known to all estimators, including the Oracle LMMSE. The Oracle LMMSE, the LMMSE mismatch filter, and A-MMSE differ only in their covariance information and in how the filter is obtained.
Reported NMSE values are computed on the $4{,}000$ test frames with one noise realization per frame, as the ratio of the summed squared error to the summed channel energy in~\eqref{eq: nmse def}. They are finite-sample estimates of the noise-averaged quantities in Appendix~\ref{app:mse}.

\begin{figure*}[t]
    \centering
    \subfigure[NMSE performance of A-MMSE variants under the Semi-Urban (SU) scenario.]{
        \includegraphics[width=0.48\textwidth]{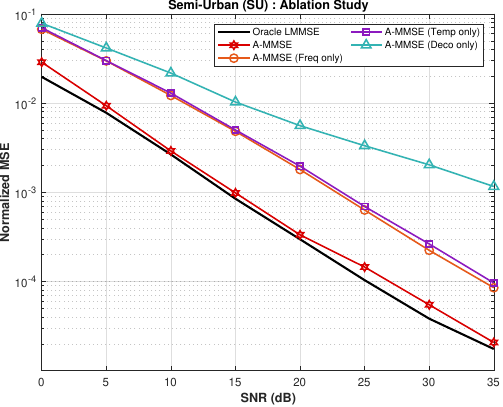}
        \label{fig: abl_su}}
    \hfill
    \subfigure[NMSE performance of A-MMSE variants under the High-Speed Rail (HSR) scenario.]{
        \includegraphics[width=0.48\textwidth]{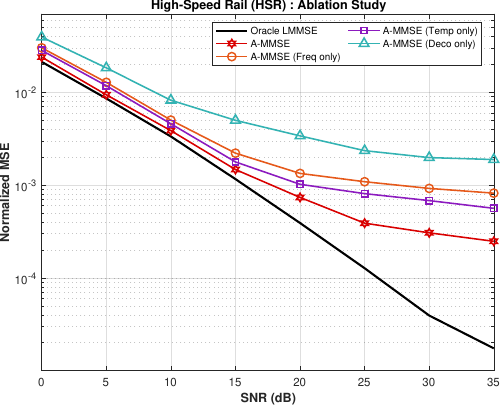}
        \label{fig: abl_hsr}}
    \caption{Ablation study: NMSE performance of A-MMSE variants under COST2100 channel models.}
    \label{fig: ammse_abl}
\end{figure*}

\subsection{Normalized MSE Comparison} \label{sec: nmse compar}

Fig.~\ref{fig:NMSE_COST} presents the NMSE as a function of SNR under both scenarios.
A-MMSE achieves lower NMSE than all practical SP-based and DNN-based baselines across every evaluated condition. The Oracle LMMSE is a lower bound rather than a competing estimator; the position of A-MMSE relative to it is quantified later in this subsection.
In the SU scenario (Fig.~\ref{fig:NMSE_SU}), characterized by rich scattering and non-stationary statistics, the A-MMSE achieves an NMSE of $2.07 \times 10^{-5}$ at an SNR of $35$ dB.
This corresponds to a reduction of approximately $73.8\%$ relative to the mismatched LMMSE ($7.91\times 10^{-5}$) and $82.2\%$ relative to Channelformer ($1.16\times 10^{-4}$).
Similar to the SU scenario, in the HSR scenario (Fig.~\ref{fig:NMSE_HSR}) characterized by high-mobility ($350$ km/h), the A-MMSE yields an NMSE of $2.48 \times 10^{-4}$ at $35$ dB.
This represents a reduction of approximately $39.4\%$ compared to the mismatched LMMSE ($4.09 \times 10^{-4}$).

The gains of A-MMSE over the mismatched LMMSE in Fig.~\ref{fig:NMSE_COST} are obtained in a non-stationary channel environment.
As described in Section~\ref{sec:sim_set}, the WSSUS assumption does not hold globally, and channel statistics vary across OFDM frames. The mismatched LMMSE relies on a fixed estimated covariance $\tilde{\mathbf{R}}$, whereas the true covariance $\mathbf{R}$ changes over time.
This is the covariance mismatch of conventional LMMSE estimation. We quantify the resulting penalty. Let $\mathbf{h} = \mathrm{vec}(\mathbf{H})$.
For a fixed filter $\mathbf{W}$, i.e., one that does not depend on the variables being averaged, let $J(\mathbf{W}) = \mathbb{E}\|\mathbf{h} - \mathbf{W}\mathbf{y}_p\|_2^2$. Let $\boldsymbol{\Sigma}_{\mathbf{yy}} = \mathbb{E}[\mathbf{y}_p\mathbf{y}_p^{\mathsf{H}}] \succ \mathbf{0}$, a second-moment matrix, and $\mathbf{W}^{*} = \mathbb{E}[\mathbf{h}\mathbf{y}_p^{\mathsf{H}}]\boldsymbol{\Sigma}_{\mathbf{yy}}^{-1}$. Then, for any fixed $\mathbf{W}$ (Appendix~\ref{app:mse}),
\begin{align}
    J(\mathbf{W}) = J(\mathbf{W}^{*}) + \underbrace{\big\|(\mathbf{W}-\mathbf{W}^{*})\boldsymbol{\Sigma}_{\mathbf{yy}}^{1/2}\big\|_F^2}_{\Delta(\mathbf{W})\,\ge\,0}.
\label{eq:mse_decomp}
\end{align}
The expectation is over the random quantities of the setting at hand, and $\mathbf{W}$ is held fixed with respect to that randomness. Two settings are used.
In the first, the expectation is over the channel and the noise.
Then $\boldsymbol{\Sigma}_{\mathbf{yy}} = \mathbf{R}_{\mathbf{h}_p\mathbf{h}_p} + \sigma^2\mathbf{I}$, $\mathbb{E}[\mathbf{h}\mathbf{y}_p^{\mathsf{H}}] = \mathbf{R}_{\mathbf{h}\mathbf{h}_p}$, and $\mathbf{W}^{*}$ has the form \eqref{eq_lmmse_structure} with the true second moments.
Let $\mathbf{W}_{\mathrm{mis}}$ be the filter \eqref{eq_lmmse_structure} with the true $\mathbf{R} = (\mathbf{R}_{\mathbf{h}\mathbf{h}_p}, \mathbf{R}_{\mathbf{h}_p\mathbf{h}_p})$ replaced by the sample second moments $\tilde{\mathbf{R}} = (\tilde{\mathbf{R}}_{\mathbf{h}\mathbf{h}_p}, \tilde{\mathbf{R}}_{\mathbf{h}_p\mathbf{h}_p})$ of Section~\ref{sec:sim_set}. Its mismatch penalty is $\Delta_{\mathrm{mis}} = \Delta(\mathbf{W}_{\mathrm{mis}})$.
Write $\Delta\mathbf{R}_1 = \tilde{\mathbf{R}}_{\mathbf{h}\mathbf{h}_p} - \mathbf{R}_{\mathbf{h}\mathbf{h}_p}$, $\Delta\mathbf{R}_2 = \tilde{\mathbf{R}}_{\mathbf{h}_p\mathbf{h}_p} - \mathbf{R}_{\mathbf{h}_p\mathbf{h}_p}$, and $\|\Delta\mathbf{R}\|_F^2 = \|\Delta\mathbf{R}_1\|_F^2 + \|\Delta\mathbf{R}_2\|_F^2$.
For fixed $\sigma^2 > 0$, the penalty admits the quadratic upper bound $\Delta_{\mathrm{mis}} \le \kappa\|\Delta\mathbf{R}\|_F^2$, where the constant $\kappa$ depends only on the true second moments and on $\sigma^2$ (Appendix~\ref{app:mse}, \eqref{eq:mis_quadratic}).
The bound requires no smallness assumption on $\Delta\mathbf{R}$. It is an upper bound only; it does not state that $\Delta_{\mathrm{mis}}$ is proportional to $\|\Delta\mathbf{R}\|_F^2$. The high-SNR limit is a separate statement.
If $\mathbf{R}_{\mathbf{h}_p\mathbf{h}_p}$ and $\tilde{\mathbf{R}}_{\mathbf{h}_p\mathbf{h}_p}$ are invertible and held fixed while $\sigma^2 \to 0$, then $\Delta_{\mathrm{mis}}$ converges to a limit that is positive exactly when the noiseless interpolators built from $\tilde{\mathbf{R}}$ and from $\mathbf{R}$ differ (Appendix~\ref{app:mse}, \eqref{eq:mis_limit}). The evaluated SNR range ends at $35$~dB, so the values reported below are measured at finite $\sigma^2$, not in this limit. At $35$~dB the penalty of the mismatched LMMSE remains large and is largest in the COST2100 HSR scenario; it is quantified below. A-MMSE does not form or invert an intermediate covariance.
It trains the Attention Transformer of Section~\ref{sec:ammse} end-to-end with the MSE loss in \eqref{eq:loss} and deploys one generated filter, held fixed. The plug-in form belongs to the same class of fixed linear filters; the two are compared on the same test frames below.

The Oracle LMMSE is a lower bound for every fixed linear filter, A-MMSE included. The Oracle LMMSE curve in Fig.~\ref{fig:NMSE_COST} is the per-frame filter $\mathbf{W}^{*}(t)$ of Section~\ref{sec:sim_set}, built from the true channel of test frame $t$.
The mismatched LMMSE and A-MMSE are single fixed filters obtained from the same $36{,}000$ training frames. For a fixed filter $\mathbf{W}$ and test frame $t$, let $J_t(\mathbf{W})$ denote the noise-averaged MSE of the estimate $\mathbf{W}\mathbf{y}_p$, and let $\boldsymbol{\Sigma}_{\mathbf{yy}}(t) = \mathbf{h}_p(t)\mathbf{h}_p(t)^{\mathsf{H}} + \sigma^2\mathbf{I}$.
This is the second moment of $\mathbf{y}_p$ conditioned on frame $t$; the conditional covariance is $\sigma^2\mathbf{I}$.
In the second setting, the expectation is over the noise of frame $t$ only. Applying \eqref{eq:mse_decomp} in this setting, for an arbitrary fixed $\mathbf{W}$, gives
\begin{align}
    J_t(\mathbf{W}) &= J_t\big(\mathbf{W}^{*}(t)\big) + \Delta_t(\mathbf{W}), \label{eq:perframe_decomp}\\
    \Delta_t(\mathbf{W}) &= \big\|\big(\mathbf{W} - \mathbf{W}^{*}(t)\big)\boldsymbol{\Sigma}_{\mathbf{yy}}(t)^{1/2}\big\|_F^2 \ge 0, \label{eq:perframe_penalty}
\end{align}
where $\mathbf{W}^{*}(t) = \mathbf{h}(t)\mathbf{h}_p(t)^{\mathsf{H}}\boldsymbol{\Sigma}_{\mathbf{yy}}(t)^{-1}$. Closed forms of $J_t(\mathbf{W})$, $\mathbf{W}^{*}(t)$, and $J_t(\mathbf{W}^{*}(t))$ are given in \eqref{eq:Jt}--\eqref{eq:Jorc_t}.
Over the test frames, the noise-averaged NMSE of any fixed $\mathbf{W}$ therefore exceeds that of the Oracle LMMSE by the average of $\Delta_t(\mathbf{W})$, normalized as in \eqref{eq: nmse def}.
Explicitly, $\mathrm{NMSE}(\mathbf{W}) = \mathrm{NMSE}(\text{Oracle LMMSE}) + \sum_t \Delta_t(\mathbf{W}) / \sum_t \|\mathbf{h}(t)\|_2^2$, with both sums over the test frames. The first term of \eqref{eq:perframe_decomp} does not depend on $\mathbf{W}$.
The mismatch penalty is therefore the only part of the MSE that the choice of a fixed filter affects. The deployed A-MMSE filter is such a fixed filter, so its excess over the Oracle LMMSE is its mismatch penalty.
Plug-in LMMSE is the comparator because it is the standard route from the same training frames to a fixed filter in the same class. In the evaluated scenarios, covariance refinement reduces the penalty but does not remove it (Remark~\ref{rem: sens}).

 At $35$~dB, the NMSE values of the Oracle LMMSE, A-MMSE, and the mismatched LMMSE are $1.75 \times 10^{-5}$, $2.07 \times 10^{-5}$, and $7.91 \times 10^{-5}$, respectively, in SU, and $1.74 \times 10^{-5}$, $2.48 \times 10^{-4}$, and $4.09 \times 10^{-4}$ in HSR.
The gap between A-MMSE and the Oracle LMMSE is thus small in SU and large in HSR.
By~\eqref{eq:Jorc_t}, the MSE of the Oracle LMMSE vanishes as $\sigma^2 \to 0$, whereas by~\eqref{eq:Jt} the MSE of a fixed filter retains an interpolation-error term that does not depend on $\sigma^2$.
In SU, where the channel varies slowly across frames, one fixed filter interpolates almost every frame well, and the NMSE of A-MMSE is $1.12\times$ that of the Oracle LMMSE at $20$~dB and $1.18\times$ at $35$~dB.
In HSR, the channel changes rapidly from frame to frame, so no fixed filter can follow every frame. This interpolation error becomes visible once the noise is small: the NMSE ratio of A-MMSE to the Oracle LMMSE grows from $1.89\times$ at $20$~dB to $14.3\times$ at $35$~dB, while the Oracle LMMSE itself is nearly the same in both scenarios at $35$~dB.
The population-optimal benchmark of Section~\ref{sec: ammse discuss} is itself a fixed linear filter.
It therefore lies above the per-frame Oracle LMMSE by its own penalty $\Delta_t(\cdot)$ on every test frame.

Regarding the RA-A-MMSE, the results indicate that the model retains estimation performance comparable to the full-rank model, even with the reduced filter rank.
In the HSR scenario, the RA-A-MMSE with $50\%$ rank achieves an NMSE of $3.01 \times 10^{-4}$ at $35$ dB.
The full-rank A-MMSE achieves an NMSE of $2.48 \times 10^{-4}$ under the same condition, illustrating the accuracy--complexity trade-off achieved by rank reduction.
Similarly, in the SU scenario, the RA-A-MMSE with $50\%$ rank shows stable estimation accuracy, yielding an NMSE of $4.84 \times 10^{-5}$ at $35$ dB.
Although the channel exhibits rich scattering, our method maintains an absolute NMSE on the order of $10^{-5}$, the same order as the full-rank A-MMSE, while reducing computational complexity.
These observations demonstrate the feasibility of the RA-A-MMSE to balance estimation fidelity and efficiency across different propagation environments.

Beyond the COST2100 model, we have also evaluated the proposed method using 3GPP-based tapped delay line (TDL) channel models, where A-MMSE consistently outperforms all baselines across all scenarios, supporting the generalizability of the method.
We have evaluated the robustness of A-MMSE to SNR mismatch as well, where the training SNR differs from the inference SNR. The results show that A-MMSE is most robust when the training SNR is chosen such that noise and clean channel components are balanced. Both sets of results are omitted here due to space constraints but are available online \cite{A-MMSE-GitHub}.

\begin{remark}\label{rem: sens} \normalfont
    We also compare our method against a sensing-aided approach~\cite{Qing2024IoTJ}, an advanced SP-based baseline.
    At $35$~dB, the sensing-aided scheme reduces NMSE by $24\%$ (SU) and $7\%$ (HSR) relative to the standard mismatched LMMSE by mitigating covariance mismatch. A-MMSE achieves a further NMSE reduction of $65\%$ and $35\%$ in SU and HSR, respectively, over this sensing-aided baseline. In the evaluated scenarios, the learned filter thus achieves lower NMSE than this covariance-refinement baseline.
\end{remark}

\subsection{Ablation Study on the Two-Stage Attention Encoder} \label{sec: abl}

The contribution of each encoder stage is assessed through three A-MMSE variants. {\it{Freq only}} retains only the Frequency Encoder, {\it{Temp only}} retains only the Temporal Encoder, and {\it{Deco only}} removes both encoder stages while preserving only the residual fully connected decoder. NMSE results under both scenarios are shown in Fig.~\ref{fig: ammse_abl}.

The {\it{Deco only}} variant produces the most severe degradation across all SNR levels, confirming that the decoder alone cannot capture the frequency-temporal correlation structure of the OFDM channel.
As shown in Fig.~\ref{fig: abl_su}, {\it{Deco only}} yields an NMSE of $1.16 \times 10^{-3}$ at $35$ dB in the SU scenario, more than $56$ times higher than the full A-MMSE ($2.07 \times 10^{-5}$).
As shown in Fig.~\ref{fig: abl_hsr}, its NMSE reaches $1.89 \times 10^{-3}$ at $35$ dB in the HSR scenario, approximately $7.6$ times higher than the full model. Both encoder stages are therefore indispensable.
Comparing the single-encoder variants reveals scenario-dependent asymmetry.
In the SU scenario, {\it{Freq only}} and {\it{Temp only}} achieve comparable NMSE of $8.52 \times 10^{-5}$ and $9.64 \times 10^{-5}$ at $35$ dB, respectively. Frequency- and temporal-domain correlations contribute roughly equally under moderate-mobility, rich-scattering conditions.
In contrast, in the HSR scenario, the Temporal Encoder is clearly more critical.
{\it{Temp only}} achieves an NMSE of $5.62 \times 10^{-4}$ at $35$ dB versus $8.19 \times 10^{-4}$ for {\it{Freq only}}, approximately $31.4\%$ lower.
The same asymmetry persists at $20$ dB. This reflects the strong Doppler effect in high-mobility environments, which amplifies the importance of temporal correlation modeling.

The full A-MMSE consistently outperforms both single-encoder variants across all SNR levels and both scenarios.
In each scenario, the gain is most pronounced relative to the weaker single-encoder variant.
At $35$ dB, the full model achieves $78.5\%$ lower NMSE than {\it{Temp only}} in SU, and $69.7\%$ lower NMSE than {\it{Freq only}} in HSR.
The Frequency and Temporal Encoders are therefore complementary rather than redundant, consistent with the Kronecker-separable covariance structure in~\eqref{eq_corr} that underpins the two-stage encoder design.
By capturing frequency- and temporal-domain correlations sequentially through dedicated stages, the architecture translates the separable structure of OFDM channel statistics into substantial estimation accuracy gains.

\begin{figure}[t]
    \centering
    \includegraphics[width=0.47\textwidth]{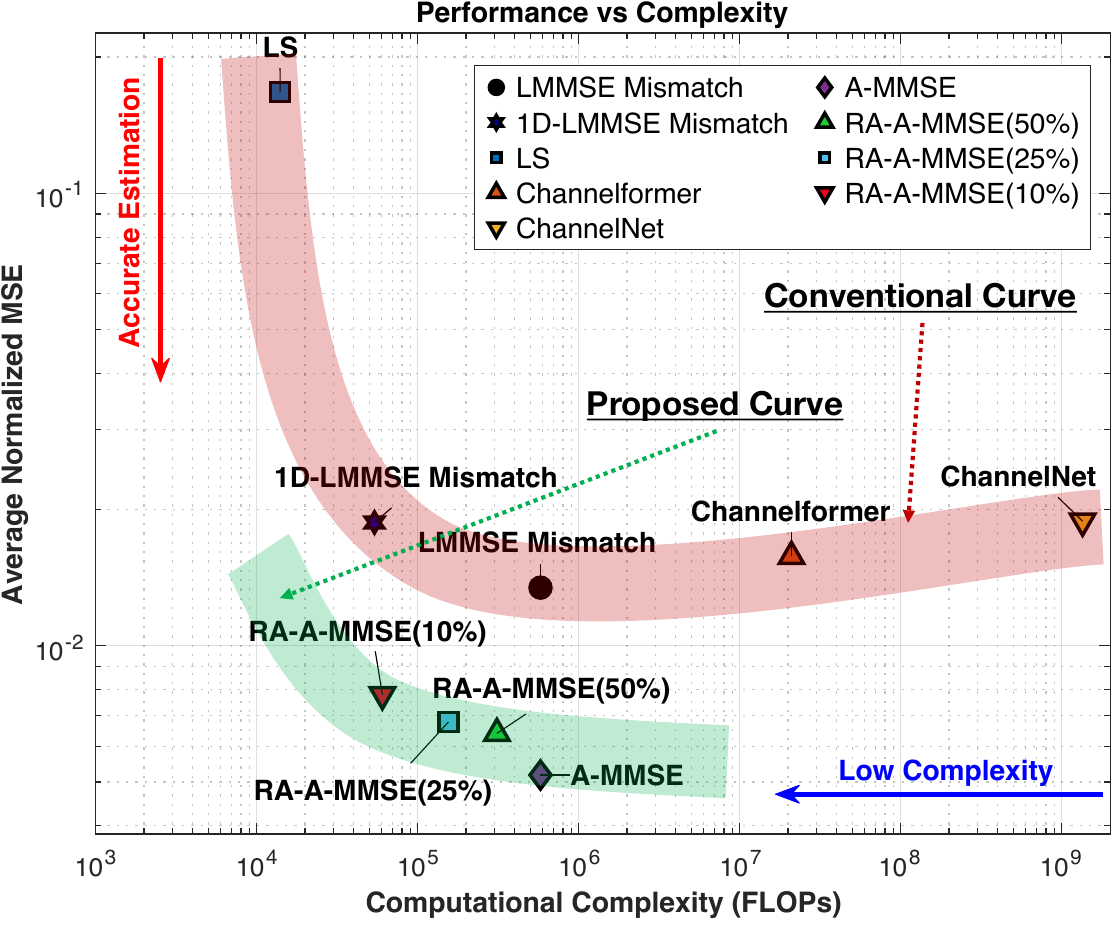}
    \caption{NMSE--complexity trade-off across channels for the proposed and conventional estimators.}
    \label{fig:COST_FLOP}
\end{figure}

\subsection{Performance-Complexity Trade-off}

In this subsection, we evaluate the computational complexity of each channel estimation method in terms of floating-point operations (FLOPs). FLOPs quantify the number of arithmetic operations required for inference (i.e., channel estimation).
Throughout this paper, FLOPs refer to the total count of real-valued arithmetic operations, including both multiplications and additions. For example, one complex multiplication $(a+jb)(c+jd)$ requires four real multiplications and two real additions, totaling six real FLOPs. All FLOP counts reported in Table~\ref{tab:model_flop} are computed consistently under this definition.
We note that training complexity is not considered in this analysis, as it is performed offline and has little impact on the real-time inference costs during deployment.

Fig.~\ref{fig:COST_FLOP} illustrates the trade-off between the average NMSE and computational complexity, averaged across all SNR levels and both HSR and SU scenarios.
We also summarize the per-inference FLOPs for all the considered methods in Table~\ref{tab:model_flop}.
As shown in Table~\ref{tab:model_flop}, A-MMSE and LMMSE share comparable inference FLOPs ($\approx 0.58$ M), as both reduce to a single matrix-vector multiplication at inference time.
The key distinction lies in estimation accuracy: A-MMSE learns the filter directly from data, whereas LMMSE relies on estimated covariance matrices that suffer from mismatch under non-stationary conditions.
As shown in Fig.~\ref{fig:COST_FLOP}, the proposed A-MMSE achieves higher estimation accuracy with substantially lower computational complexity.
Specifically, it yields around $64.5\%$ lower average NMSE and $97.2\%$ fewer FLOPs compared to the average of Channelformer.
As further summarized in Table~\ref{tab:model_flop}, A-MMSE achieves a per-inference runtime of $\sim 0.3739$ ms, which is over $55\times$ and $70\times$ faster than Channelformer and ChannelNet, respectively, while maintaining a comparable memory footprint of $\sim 0.562$ MB relative to Channelformer.
Notably, this runtime is on par with the classical LMMSE estimator, confirming that A-MMSE achieves DNN-level estimation accuracy at SP-level inference speed.
The RA-A-MMSE further reduces memory consumption to $\sim 0.181$ MB and achieves a runtime of $\sim 0.327$ ms, where both values are averaged over the $50\%$, $25\%$, and $10\%$ rank configurations, making it particularly well-suited for deployment on memory-constrained receivers.
A detailed analysis of memory and runtime across all considered methods is omitted here due to page limitations, and is deferred to an extended version of this work.

The rank-adaptive extension further optimizes this efficiency. The RA-A-MMSE with $10\%$ rank retains high accuracy despite the reduced rank, achieving approximately $50.5\%$ lower average NMSE compared to Channelformer.
This demonstrates that the RA-A-MMSE effectively exploits the estimator's low-rank structure, offering a scalable solution that outperforms DNN baselines with reduced computational overhead.
As discussed, these benefits arise from the fact that A-MMSE relies solely on computationally efficient linear operations during channel estimation.
Consequently, the A-MMSE and RA-A-MMSE frameworks advance the performance--complexity trade-off, achieving improved estimation accuracy at significantly reduced computational cost.
This establishes a practical benchmark for channel estimation by enhancing the trade-off between accuracy and efficiency.

\begin{table}[t]
\centering
\caption{FLOPs, Memory and Runtime Comparison for Channel Estimation ($L=72$)}
\label{tab:model_flop}
\renewcommand{\arraystretch}{1.0}
\begin{tabular}{l|c|c|c}
\noalign{\hrule height 1.5pt}
\textbf{Method} & \textbf{FLOPs} & \textbf{Memory} & \textbf{Runtime} \\
\noalign{\hrule height 1.0pt}

\textbf{LMMSE} & $\approx 0.58$ M & -- & $\approx 0.37$ ms \\ \hline
\textbf{1D-LMMSE} & $\approx 54$ K & -- & $\approx 0.077$ ms \\ \hline
\textbf{LS} & $\approx 14$ K & -- & $\approx 0.038$ ms \\ \hline
\textbf{ChannelNet} & $\approx 1.35$ G & 1.760 MB & 26.281 ms \\ \hline
\textbf{Channelformer} & $\approx 21$ M & 0.610 MB & 20.766 ms \\ \hline
\textbf{A-MMSE} & $\approx 0.58$ M & 0.562 MB & 0.3739 ms \\ \hline
\textbf{RA-A-MMSE} & $\approx 8640r,\; r=\text{rank}$ & 0.181 MB & 0.327 ms \\

\noalign{\hrule height 1.5pt}
\end{tabular}
\end{table}

\section{Conclusion} \label{sec:con}
In this paper, we proposed the A-MMSE method for OFDM channel estimation.
Our key idea is to integrate the classical SP-based framework with the learning capability of the DNN-based approach.
Specifically, the proposed A-MMSE learns the linear filtering matrix via the Attention Transformer and applies it for channel estimation, achieving both high estimation accuracy and substantial computational efficiency.
In particular, to enhance the learning of the linear filtering matrix, we developed a novel two-stage Attention encoder that effectively captures the frequency and temporal correlation structures of the OFDM channel.
Building on this architecture, we further introduced a rank-adaptive extension of A-MMSE, referred to as the RA-A-MMSE.
The RA-A-MMSE is distinguished by adjusting the rank of the linear filtering matrix at deployment time, thereby controlling the computational overhead associated with channel estimation.
Simulations with COST2100 channel models confirmed that A-MMSE outperforms conventional SP-based and DNN-based estimators. The RA-A-MMSE maintains comparable accuracy while substantially reducing computational complexity.

The proposed frameworks stand out in these aspects, namely i) architectural simplicity at inference through a single linear operation, ii) effective capture of the intrinsic channel correlation structure via the two-stage Attention encoder, and iii) principled integration of classical signal processing with deep learning.
Together, these aspects position A-MMSE as a scalable and AI-native solution well aligned with emerging 6G physical-layer requirements.

Future work may extend this framework to scenarios where ground-truth channel information is unavailable during training or to MIMO-OFDM channel acquisition \cite{kim:twc:25}.
Exploring reinforcement learning paradigms that leverage end-to-end performance metrics represents a promising direction.
Furthermore, jointly accounting for additional deployment factors in the rank selection represents a promising extension.

\appendices
\section{MSE Decomposition, Covariance Mismatch, and Oracle Bounds}
\label{app:mse}

Throughout, a filter $\mathbf{W}$ is called fixed if it does not depend on the random variables over which the stated expectation is taken.
The two instances below differ only in what is averaged: the channel and the noise, or the noise of one frame. A learned filter is treated in the same way: the outcome of training is held fixed and then evaluated.

\textit{Lemma~1:} Let $\mathbf{h}$ and $\mathbf{y}_p$ be random vectors with finite second moments and $\boldsymbol{\Sigma}_{\mathbf{yy}} = \mathbb{E}[\mathbf{y}_p\mathbf{y}_p^{\mathsf{H}}] \succ \mathbf{0}$. Let $\mathbf{C} = \mathbb{E}[\mathbf{h}\mathbf{y}_p^{\mathsf{H}}]$, $\mathbf{W}^{*} = \mathbf{C}\boldsymbol{\Sigma}_{\mathbf{yy}}^{-1}$, and $J(\mathbf{W}) = \mathbb{E}\|\mathbf{h} - \mathbf{W}\mathbf{y}_p\|_2^2$. Then~\eqref{eq:mse_decomp} holds for every fixed $\mathbf{W}$, i.e., every $\mathbf{W}$ that does not depend on $\mathbf{h}$ and $\mathbf{y}_p$, with $\Delta(\mathbf{W}) = \|(\mathbf{W} - \mathbf{W}^{*})\boldsymbol{\Sigma}_{\mathbf{yy}}^{1/2}\|_F^2$.

\textit{Proof:} Since $\mathbf{C}^{\mathsf{H}} = \boldsymbol{\Sigma}_{\mathbf{yy}}\mathbf{W}^{*\mathsf{H}}$ and $\mathbf{W}$ is fixed, expanding the square gives
\begin{align}
    J(\mathbf{W}) = \mathbb{E}\|\mathbf{h}\|_2^2 - 2\,\Re\,\mathrm{tr}\big(\mathbf{W}\boldsymbol{\Sigma}_{\mathbf{yy}}\mathbf{W}^{*\mathsf{H}}\big) + \mathrm{tr}\big(\mathbf{W}\boldsymbol{\Sigma}_{\mathbf{yy}}\mathbf{W}^{\mathsf{H}}\big).
\label{eq:lemma_expand}
\end{align}
Substitute $\mathbf{W} = \mathbf{W}^{*} + \mathbf{E}$. The terms linear in $\mathbf{E}$ cancel, and~\eqref{eq:lemma_expand} becomes
\begin{align}
    J(\mathbf{W}) = \mathbb{E}\|\mathbf{h}\|_2^2 - \mathrm{tr}\big(\mathbf{W}^{*}\boldsymbol{\Sigma}_{\mathbf{yy}}\mathbf{W}^{*\mathsf{H}}\big) + \big\|\mathbf{E}\boldsymbol{\Sigma}_{\mathbf{yy}}^{1/2}\big\|_F^2.
\label{eq:lemma_square}
\end{align}
The first two terms equal $J(\mathbf{W}^{*})$ (set $\mathbf{E} = \mathbf{0}$), and the last term is $\Delta(\mathbf{W})$. The proof uses neither Gaussianity nor a noise model. \hfill$\blacksquare$

\textit{Covariance mismatch:} Let $\mathbf{y}_p = \mathbf{h}_p + \mathbf{z}_p$, with the expectation over the channel and the noise.
Under the noise model of \eqref{eq_lmmse_structure}, the noise is zero-mean, has second moment $\sigma^2\mathbf{I}$, and is uncorrelated with the channel. Here, $\mathbf{R}_{\mathbf{h}\mathbf{h}_p} := \mathbb{E}[\mathbf{h}\mathbf{h}_p^{\mathsf{H}}]$ and $\mathbf{R}_{\mathbf{h}_p\mathbf{h}_p} := \mathbb{E}[\mathbf{h}_p\mathbf{h}_p^{\mathsf{H}}]$ denote the uncentered second moments. Then $\boldsymbol{\Sigma}_{\mathbf{yy}} = \mathbf{R}_{\mathbf{h}_p\mathbf{h}_p} + \sigma^2\mathbf{I}$, $\mathbf{C} = \mathbf{R}_{\mathbf{h}\mathbf{h}_p}$, and $\mathbf{W}^{*}$ has the form \eqref{eq_lmmse_structure} with the true second moments.
Write $\tilde{\mathbf{R}}_{\mathbf{h}\mathbf{h}_p} = \mathbf{R}_{\mathbf{h}\mathbf{h}_p} + \Delta\mathbf{R}_1$, $\tilde{\mathbf{R}}_{\mathbf{h}_p\mathbf{h}_p} = \mathbf{R}_{\mathbf{h}_p\mathbf{h}_p} + \Delta\mathbf{R}_2$, and $\tilde{\boldsymbol{\Sigma}} = \boldsymbol{\Sigma}_{\mathbf{yy}} + \Delta\mathbf{R}_2$, so that $\mathbf{W}_{\mathrm{mis}} = \tilde{\mathbf{R}}_{\mathbf{h}\mathbf{h}_p}\tilde{\boldsymbol{\Sigma}}^{-1}$ and $\Delta_{\mathrm{mis}} = \Delta(\mathbf{W}_{\mathrm{mis}})$. The sample second moments are computed from the training frames and are held fixed here. Since $\tilde{\boldsymbol{\Sigma}}^{-1} - \boldsymbol{\Sigma}_{\mathbf{yy}}^{-1} = -\boldsymbol{\Sigma}_{\mathbf{yy}}^{-1}\Delta\mathbf{R}_2\tilde{\boldsymbol{\Sigma}}^{-1}$,
\begin{align}
    \mathbf{W}_{\mathrm{mis}} - \mathbf{W}^{*} = \big(\Delta\mathbf{R}_1 - \mathbf{W}^{*}\Delta\mathbf{R}_2\big)\tilde{\boldsymbol{\Sigma}}^{-1}
\label{eq:mis_diff}
\end{align}
holds exactly. Since $\tilde{\mathbf{R}}_{\mathbf{h}_p\mathbf{h}_p} \succeq \mathbf{0}$, $\|\tilde{\boldsymbol{\Sigma}}^{-1}\|_2 \le \sigma^{-2}$. Inserting~\eqref{eq:mis_diff} into~\eqref{eq:mse_decomp} and bounding the norms of the products gives
\begin{align}
    \Delta_{\mathrm{mis}} \le \sigma^{-4}\|\boldsymbol{\Sigma}_{\mathbf{yy}}\|_2\big(\|\Delta\mathbf{R}_1\|_F + \|\mathbf{W}^{*}\|_2\|\Delta\mathbf{R}_2\|_F\big)^2,
\label{eq:mis_bound}
\end{align}
with no smallness assumption on $\Delta\mathbf{R}_1$ and $\Delta\mathbf{R}_2$. Since $(a + cb)^2 \le 2\max\{1, c^2\}(a^2 + b^2)$,~\eqref{eq:mis_bound} gives
\begin{align}
    \Delta_{\mathrm{mis}} \le \kappa\|\Delta\mathbf{R}\|_F^2,\qquad \kappa = 2\sigma^{-4}\|\boldsymbol{\Sigma}_{\mathbf{yy}}\|_2\max\{1, \|\mathbf{W}^{*}\|_2^2\},
\label{eq:mis_quadratic}
\end{align}
with $\|\Delta\mathbf{R}\|_F^2 = \|\Delta\mathbf{R}_1\|_F^2 + \|\Delta\mathbf{R}_2\|_F^2$. The constant $\kappa$ depends only on the true second moments and on $\sigma^2 > 0$. This is the quadratic upper bound used in Section~\ref{sec: nmse compar}. It bounds $\Delta_{\mathrm{mis}}$ from above; it does not assert that $\Delta_{\mathrm{mis}}$ scales in proportion to $\|\Delta\mathbf{R}\|_F^2$. The high-SNR limit is a separate statement. Assume $\mathbf{R}_{\mathbf{h}_p\mathbf{h}_p} \succ \mathbf{0}$ and $\tilde{\mathbf{R}}_{\mathbf{h}_p\mathbf{h}_p} \succ \mathbf{0}$, and hold these second moments fixed while $\sigma^2 \to 0$. Then
\begin{align}
\Delta_{\mathrm{mis}} \to \Big\|\Big(\tilde{\mathbf{R}}_{\mathbf{h}\mathbf{h}_p}\tilde{\mathbf{R}}_{\mathbf{h}_p\mathbf{h}_p}^{-1} - \mathbf{R}_{\mathbf{h}\mathbf{h}_p}\mathbf{R}_{\mathbf{h}_p\mathbf{h}_p}^{-1}\Big)\mathbf{R}_{\mathbf{h}_p\mathbf{h}_p}^{1/2}\Big\|_F^2.
\label{eq:mis_limit}
\end{align}
The limit is positive if and only if the two noiseless interpolators differ, i.e., the $\sigma^2 = 0$ forms of~\eqref{eq_lmmse_structure} built from $\tilde{\mathbf{R}}$ and from $\mathbf{R}$.

\textit{Per-frame oracle:} Fix a test frame $t$, so that $\mathbf{h}(t)$ and its pilot entries $\mathbf{h}_p(t)$ are deterministic. Let $\mathbf{y}_p = \mathbf{h}_p(t) + \mathbf{z}_p$ with $\mathbb{E}[\mathbf{z}_p] = \mathbf{0}$ and $\mathbb{E}[\mathbf{z}_p\mathbf{z}_p^{\mathsf{H}}] = \sigma^2\mathbf{I}$, the expectation being over $\mathbf{z}_p$ only. For a filter $\mathbf{W}$ fixed with respect to $\mathbf{z}_p$, the cross term has zero mean, and
\begin{align}
    J_t(\mathbf{W}) = \|\mathbf{h}(t) - \mathbf{W}\mathbf{h}_p(t)\|_2^2 + \sigma^2\|\mathbf{W}\|_F^2.
\label{eq:Jt}
\end{align}
Here $\boldsymbol{\Sigma}_{\mathbf{yy}} = \boldsymbol{\Sigma}_{\mathbf{yy}}(t)$ and $\mathbf{C} = \mathbf{h}(t)\mathbf{h}_p(t)^{\mathsf{H}}$. Since $\boldsymbol{\Sigma}_{\mathbf{yy}}(t)$ is a rank-one update of $\sigma^2\mathbf{I}$, $\mathbf{h}_p(t)^{\mathsf{H}}\boldsymbol{\Sigma}_{\mathbf{yy}}(t)^{-1} = \mathbf{h}_p(t)^{\mathsf{H}}/(\|\mathbf{h}_p(t)\|_2^2 + \sigma^2)$, and $\mathbf{W}^{*} = \mathbf{C}\boldsymbol{\Sigma}_{\mathbf{yy}}(t)^{-1}$ reduces to
\begin{align}
    \mathbf{W}^{*}(t) = \frac{\mathbf{h}(t)\mathbf{h}_p(t)^{\mathsf{H}}}{\|\mathbf{h}_p(t)\|_2^2 + \sigma^2}.
\label{eq:Wstar_t}
\end{align}
Substituting~\eqref{eq:Wstar_t} into~\eqref{eq:Jt} gives
\begin{align}
    J_t\big(\mathbf{W}^{*}(t)\big) = \frac{\sigma^2\|\mathbf{h}(t)\|_2^2}{\|\mathbf{h}_p(t)\|_2^2 + \sigma^2}.
\label{eq:Jorc_t}
\end{align}
Lemma~1 then gives~\eqref{eq:perframe_decomp}--\eqref{eq:perframe_penalty}. The filter $\mathbf{W}^{*}(t)$ uses the true channel of frame $t$. It is the reference for the class of estimators $\hat{\mathbf{h}} = \mathbf{W}\mathbf{y}_p$ with $\mathbf{W}$ fixed with respect to the noise, and it is not realizable by a receiver. By~\eqref{eq:Jorc_t}, $J_t(\mathbf{W}^{*}(t)) \to 0$ as $\sigma^2 \to 0$ on every frame with $\|\mathbf{h}_p(t)\|_2^2 > 0$. Summing~\eqref{eq:perframe_decomp} over the test frames and dividing by $\sum_t \|\mathbf{h}(t)\|_2^2$ gives the NMSE identity of Section~\ref{sec: nmse compar}; there,~\eqref{eq: nmse def} is evaluated on the test frames as a ratio of sums.

\bibliographystyle{IEEEtran}
\bibliography{ammse_ref}

\begin{thebibliography}{10}
\providecommand{\url}[1]{#1}
\csname url@samestyle\endcsname
\providecommand{\newblock}{\relax}
\providecommand{\bibinfo}[2]{#2}
\providecommand{\BIBentrySTDinterwordspacing}{\spaceskip=0pt\relax}
\providecommand{\BIBentryALTinterwordstretchfactor}{4}
\providecommand{\BIBentryALTinterwordspacing}{\spaceskip=\fontdimen2\font plus
\BIBentryALTinterwordstretchfactor\fontdimen3\font minus
  \fontdimen4\font\relax}
\providecommand{\BIBforeignlanguage}[2]{{%
\expandafter\ifx\csname l@#1\endcsname\relax
\typeout{** WARNING: IEEEtran.bst: No hyphenation pattern has been}%
\typeout{** loaded for the language `#1'. Using the pattern for}%
\typeout{** the default language instead.}%
\else
\language=\csname l@#1\endcsname
\fi
#2}}
\providecommand{\BIBdecl}{\relax}
\BIBdecl

\bibitem{shafi:jsac:17}
M.~Shafi, A.~F. Molisch, P.~J. Smith, T.~Haustein, P.~Zhu, P.~De~Silva,
  F.~Tufvesson, A.~Benjebbour, and G.~Wunder, ``{5G}: {A} tutorial overview of
  standards, trials, challenges, deployment, and practice,'' \emph{IEEE J. Sel.
  Areas Commun.}, vol.~35, no.~6, pp. 1201--1221, 2017.

\bibitem{guan:jsac:17}
P.~Guan, D.~Wu, T.~Tian, J.~Zhou, X.~Zhang, L.~Gu, A.~Benjebbour, M.~Iwabuchi,
  and Y.~Kishiyama, ``{5G} field trials: {OFDM}-based waveforms and mixed
  numerologies,'' \emph{IEEE J. Sel. Areas Commun.}, vol.~35, no.~6, pp.
  1234--1243, 2017.

\bibitem{liy:twc:21}
S.~D. Liyanaarachchi, T.~Riihonen, C.~B. Barneto, and M.~Valkama, ``Optimized
  waveforms for {5G–6G} communication with sensing: {Theory}, simulations and
  experiments,'' \emph{IEEE Trans. Wireless Commun.}, vol.~20, no.~12, pp.
  8301--8315, 2021.

\bibitem{edfors:tcom:98}
O.~Edfors, M.~Sandell, J.-J. van~de Beek, S.~Wilson, and P.~Borjesson, ``{OFDM}
  channel estimation by singular value decomposition,'' \emph{IEEE Trans.
  Commun.}, vol.~46, no.~7, pp. 931--939, 1998.

\bibitem{li:tvt:00}
Y.~Li, ``Pilot-symbol-aided channel estimation for {OFDM} in wireless
  systems,'' \emph{IEEE Trans. Veh. Technol.}, vol.~49, no.~4, pp. 1207--1215,
  2000.

\bibitem{liu:survey:14}
Y.~Liu, Z.~Tan, H.~Hu, L.~J. Cimini, and G.~Y. Li, ``Channel estimation for
  {OFDM},'' \emph{IEEE Commun. Surveys \& Tutorials}, vol.~16, no.~4, pp.
  1891--1908, 2014.

\bibitem{soltani:commlett:19}
M.~Soltani, V.~Pourahmadi, A.~Mirzaei, and H.~Sheikhzadeh, ``Deep
  learning-based channel estimation,'' \emph{IEEE Commun. Lett.}, vol.~23,
  no.~4, pp. 652--655, 2019.

\bibitem{li:wcl:20}
L.~Li, H.~Chen, H.-H. Chang, and L.~Liu, ``Deep residual learning meets {OFDM}
  channel estimation,'' \emph{IEEE Wireless Commun. Lett.}, vol.~9, no.~5, pp.
  615--618, 2020.

\bibitem{luan:wsa:21}
D.~Luan and J.~Thompson, ``Low complexity channel estimation with neural
  network solutions,'' in \emph{Proc. 25th International ITG Workshop on Smart
  Antennas (WSA)}, 2021, pp. 1--6.

\bibitem{balevi:jsac:21}
E.~Balevi, A.~Doshi, A.~Jalal, A.~Dimakis, and J.~G. Andrews, ``High
  dimensional channel estimation using deep generative networks,'' \emph{IEEE
  J. Sel. Areas Commun.}, vol.~39, no.~1, pp. 18--30, 2021.

\bibitem{vaswani2017attention}
A.~Vaswani, N.~Shazeer, N.~Parmar, J.~Uszkoreit, L.~Jones, A.~N. Gomez,
  {\L}.~Kaiser, and I.~Polosukhin, ``Attention is all you need,'' in
  \emph{Proc. Adv. Neural Inf. Process. Syst. (NeurIPS)}, vol.~30, 2017, pp.
  5998--6008.

\bibitem{li2023wireless}
W.~Zheng, R.~Wang, Y.~Yuan, J.~Li, B.~He, and F.~Lin, ``Wireless channel
  estimation based on transformer and super-resolution,'' in \emph{Proceedings
  of the 2024 3rd Asia Conference on Algorithms, Computing and Machine Learning
  (CACML '24)}.\hskip 1em plus 0.5em minus 0.4em\relax New York, NY, USA:
  Association for Computing Machinery, 2024, pp. 419--423.

\bibitem{luan2023channelformer}
D.~Luan and J.~S. Thompson, ``Channelformer: {Attention} based neural solution
  for wireless channel estimation and effective online training,'' \emph{IEEE
  Trans. Wireless Commun.}, vol.~22, no.~10, pp. 6562--6577, 2023.

\bibitem{nair2010rectified}
V.~Nair and G.~E. Hinton, ``Rectified linear units improve restricted
  {Boltzmann} machines,'' in \emph{Proceedings of the International Conference
  on Machine Learning (ICML)}, 2010, pp. 807--814.

\bibitem{lecun2012efficient}
Y.~LeCun, L.~Bottou, G.~B. Orr, and K.~R. Müller, ``Efficient backprop,'' in
  \emph{Neural Networks: Tricks of the Trade}.\hskip 1em plus 0.5em minus
  0.4em\relax Springer, 2012, pp. 9--48.

\bibitem{clevert2015fast}
D.~A. Clevert, T.~Unterthiner, and S.~Hochreiter, ``Fast and accurate deep
  network learning by exponential linear units {(ELUs)},'' in \emph{Proc. Int.
  Conf. Learn. Represent. (ICLR)}, 2016.

\bibitem{scherer2010evaluation}
D.~Scherer, A.~Müller, and S.~Behnke, ``Evaluation of pooling operations in
  convolutional architectures for object recognition,'' in \emph{Artificial
  Neural Networks–ICANN 2010}.\hskip 1em plus 0.5em minus 0.4em\relax
  Springer, 2010, pp. 92--101.

\bibitem{greenberg:neurips:23}
I.~Greenberg, N.~Yannay, and S.~Mannor, ``Optimization or architecture: {How}
  to hack {Kalman} filtering,'' in \emph{Proc. of the Int. Conf. on Neural
  Information Processing Systems (NeurIPS)}, 2023.

\bibitem{ngu:wcommagg:23}
L.~V. Nguyen, N.~T. Nguyen, N.~H. Tran, M.~Juntti, A.~L. Swindlehurst, and
  D.~H.~N. Nguyen, ``Leveraging deep neural networks for massive {MIMO} data
  detection,'' \emph{IEEE Wireless Commun.}, vol.~30, no.~1, pp. 174--180,
  2023.

\bibitem{mbdnn:book:23}
N.~Shlezinger and Y.~C. Eldar, ``Model-based deep learning,'' \emph{Foundations
  and Trends® in Signal Processing}, vol.~17, no.~4, pp. 291--416, 2023.

\bibitem{kalmannet:tsp:22}
G.~Revach, N.~Shlezinger, X.~Ni, A.~L. Escoriza, R.~J.~G. van Sloun, and Y.~C.
  Eldar, ``{KalmanNet}: {Neural} network aided {Kalman} filtering for partially
  known dynamics,'' \emph{IEEE Trans. Signal Process.}, vol.~70, pp.
  1532--1547, 2022.

\bibitem{welch:kalman}
G.~Welch and G.~Bishop, ``An introduction to the {Kalman} filter,'' University
  of North Carolina at Chapel Hill, USA, Tech. Rep., 1995.

\bibitem{lstm:97}
S.~Hochreiter and J.~Schmidhuber, ``Long short-term memory,'' \emph{Neural
  Computation}, vol.~9, no.~8, pp. 1735--1780, 1997.

\bibitem{cho:gru:14}
K.~Cho, B.~van Merri{\"e}nboer, C.~Gulcehre, D.~Bahdanau, F.~Bougares,
  H.~Schwenk, and Y.~Bengio, ``Learning phrase representations using {RNN}
  encoder{--}decoder for statistical machine translation,'' in \emph{Proc. the
  Conference on Empirical Methods in Natural Language Process. ({EMNLP})}, Oct.
  2014, pp. 1724--1734.

\bibitem{choi:tvt:23}
G.~Choi, J.~Park, N.~Shlezinger, Y.~C. Eldar, and N.~Lee, ``{Split-KalmanNet}:
  {A} robust model-based deep learning approach for state estimation,''
  \emph{IEEE Trans. Veh. Technol.}, vol.~72, no.~9, pp. 12\,326--12\,331, 2023.

\bibitem{shen2025kalmanformer}
S.~Shen, J.~Chen, G.~Yu, Z.~Zhai, and P.~Han, ``{KalmanFormer}: Using
  transformer to model the {Kalman} gain in {Kalman} filters,'' \emph{Frontiers
  in Neurorobotics}, vol.~18, p. 1460255, 2025.

\bibitem{aikalman:25}
N.~Shlezinger, G.~Revach, A.~Ghosh, S.~Chatterjee, S.~Tang, T.~Imbiriba,
  J.~Dunik, O.~Straka, P.~Closas, and Y.~C. Eldar, ``Artificial
  intelligence-aided {Kalman} filters: {AI}-augmented designs for {Kalman}-type
  algorithms,'' \emph{IEEE Signal Process. Mag.}, vol.~42, no.~3, pp. 52--76,
  2025.

\bibitem{jungBussgang}
\BIBentryALTinterwordspacing
C.~Jung, T.~Ha, H.~Kim, and J.~Park, ``State estimation with 1-bit observations
  and imperfect models: {Bussgang} meets {Kalman} in neural networks,'' 2025.
  [Online]. Available: \url{https://arxiv.org/abs/2507.17284}
\BIBentrySTDinterwordspacing

\bibitem{xu:twc:25}
Q.~Xu, J.~Sun, and Z.~Xu, ``Efficient {MU-MIMO} beamforming based on
  majorization-minimization and deep unfolding,'' \emph{IEEE Trans. Wireless
  Commun.}, vol.~24, no.~5, pp. 3949--3963, 2025.

\bibitem{khani:twc:20}
M.~Khani, M.~Alizadeh, J.~Hoydis, and P.~Fleming, ``Adaptive neural signal
  detection for massive {MIMO},'' \emph{IEEE Trans. Wireless Commun.}, vol.~19,
  no.~8, pp. 5635--5648, 2020.

\bibitem{Sun2025twc}
H.~Sun, L.~Zhu, W.~Mei, and R.~Zhang, ``Power-measurement-based channel
  autocorrelation estimation for {IRS}-assisted wideband communications,''
  \emph{IEEE Transactions on Wireless Communications}, vol.~24, no.~6, pp.
  4647--4662, 2025.

\bibitem{Sun2024twc}
------, ``Power measurement-based channel estimation for {IRS}-enhanced
  wireless coverage,'' \emph{IEEE Transactions on Wireless Communications},
  vol.~23, no.~12, pp. 19\,183--19\,198, 2024.

\bibitem{Liu2025CL}
Y.~Liu, W.~Mei, H.~Sun, D.~Wang, and Z.~Chen, ``Power-measurement-based channel
  estimation for beyond diagonal {RIS},'' \emph{IEEE Communications Letters},
  vol.~29, no.~11, pp. 2666--2670, 2025.

\bibitem{Yoojin20152DLMMSE}
Y.~Choi, J.~H. Bae, and J.~Lee, ``Low-complexity {2D} {LMMSE} channel
  estimation for {OFDM} systems,'' in \emph{Proc. IEEE Veh. Technol. Conf.},
  2015, pp. 1--5.

\bibitem{COST2100}
L.~Liu, C.~Oestges, J.~Poutanen, K.~Haneda, P.~Vainikainen, F.~Quitin,
  F.~Tufvesson, and P.~{De Doncker}, ``The {COST} 2100 {MIMO} channel model,''
  \emph{IEEE Wireless Commun.}, vol.~19, no.~6, pp. 92--99, 2012.

\bibitem{coleri:tbc:02}
S.~Coleri, M.~Ergen, A.~Puri, and A.~Bahai, ``Channel estimation techniques
  based on pilot arrangement in {OFDM} systems,'' \emph{IEEE Trans.
  Broadcast.}, vol.~48, no.~3, pp. 223--229, 2002.

\bibitem{vanloan1993approximation}
C.~F. {Van Loan} and N.~Pitsianis, ``Approximation with {Kronecker} products,''
  in \emph{Linear Algebra for Large Scale and Real-Time Applications}, ser.
  NATO ASI Series, M.~S. Moonen, G.~H. Golub, and B.~L.~R. {De Moor},
  Eds.\hskip 1em plus 0.5em minus 0.4em\relax Dordrecht, The Netherlands:
  Springer, 1993, vol. 232, pp. 293--314.

\bibitem{rel:conv:att:2020}
\BIBentryALTinterwordspacing
J.-B. Cordonnier, A.~Loukas, and M.~Jaggi, ``On the relationship between
  self-attention and convolutional layers,'' in \emph{International Conference
  on Learning Representations (ICLR)}, 2020. [Online]. Available:
  \url{https://arxiv.org/abs/1911.03584}
\BIBentrySTDinterwordspacing

\bibitem{kingma2014adam}
D.~P. Kingma and J.~Ba, ``Adam: A method for stochastic optimization,'' in
  \emph{Proc. Int. Conf. Learn. Represent. (ICLR)}, 2015.

\bibitem{3GPP-TS-38.211}
{3rd Generation Partnership Project (3GPP)}, ``{NR}; physical channels and
  modulation,'' 3GPP, Tech. Spec. TS 38.211, April 2025, release 18.

\bibitem{A-MMSE-GitHub}
T.~Ha, \url{https://github.com/TaeJun1999/Attention-aided-MMSE}.

\bibitem{Qing2024IoTJ}
C.~Qing, W.~Hu, Z.~Liu, G.~Ling, X.~Cai, and P.~Du, ``Sensing-aided channel
  estimation in {OFDM} systems by leveraging communication echoes,'' \emph{IEEE
  Internet of Things J.}, vol.~11, no.~23, pp. 38\,023--38\,039, 2024.

\bibitem{kim:twc:25}
N.~Kim, I.~P. Roberts, and J.~Park, ``Splitting messages in the
  dark--{Rate}-splitting multiple access for {FDD} massive {MIMO} without {CSI}
  feedback,'' \emph{IEEE Trans. Wireless Commun.}, vol.~24, no.~4, pp.
  3320--3332, 2025.

\end{thebibliography}

\end{document}